\documentclass[manuscript,screen]{acmart}

\usepackage{graphicx} 
\usepackage{subcaption} 
\usepackage{amsmath}
\usepackage{amsfonts}
\usepackage{amsthm}
\usepackage{algorithmic}
\usepackage{textcomp}
\usepackage{xcolor}
\usepackage{bm}
\usepackage{makecell}
\usepackage{tabu}
\usepackage[ruled,boxed,linesnumbered]{algorithm2e}
\usepackage{multirow}
\usepackage{enumitem}
\usepackage{url}
\usepackage{colortbl}
\usepackage{multicol}
\usepackage{xspace}
\usepackage{bbding}
\usepackage{pifont}
\usepackage{tabularx}
\usepackage{array} 
\usepackage{forest}
\usepackage{adjustbox}
\usepackage{longtable}
\usepackage[most]{tcolorbox}

\newcolumntype{Y}{>{\centering\arraybackslash}X}

\newtheorem{definition}{{Definition}}

\AtBeginDocument{%
  }

\setcopyright{acmlicensed}
\copyrightyear{2025}
\acmYear{2025}
\acmDOI{XXXXXXX.XXXXXXX}

\acmConference[]{ACM COMPUTING SURVEYS}{
  2025}{Woodstock, NY}
\acmISBN{978-1-4503-XXXX-X/18/06}

\begin{document}

\title{AI-Empowered Catalyst Discovery: A Survey from Classical Machine Learning Approaches to Large Language Models}
\author{Yuanyuan Xu}
\affiliation{%
  \institution{School of Computer Science and Engineering, The University of New South Wales}
  \city{Sydney}
  \country{Australia}}
\email{yuanyuan.xu@unsw.edu.au}

\author{Hanchen Wang}
\affiliation{%
  \institution{Australian Artificial Intelligence Institute, University of Technology Sydney}
  \city{Sydney}
  \country{Australia}}
\email{hanchen.wang@uts.edu.au}

\author{Wenjie Zhang}
\affiliation{%
  \institution{School of Computer Science and Engineering, The University of New South Wales}
  \city{Sydney}
  \country{Australia}}
\email{wenjie.zhang@unsw.edu.au}

\author{Lexing Xie}
\affiliation{%
  \institution{School of Computing, Australian National University}
  \city{Canberra}
  \country{Australia}}
\email{lexing.xie@anu.edu.au}

\author{Yin Chen}
\affiliation{%
  \institution{Faculty of Engineering and Information Technology, University of Technology Sydney}
  \city{Sydney}
  \country{Australia}}
\email{yin.chen@student.uts.edu.au}

\author{Flora Salim}
\affiliation{%
  \institution{School of Computer Science and Engineering, The University of New South Wales}
  \city{Sydney}
  \country{Australia}}
\email{flora.salim@unsw.edu.au}

\author{Ying Zhang}
\affiliation{%
  \institution{Australian Artificial Intelligence Institute, University of Technology Sydney}
  \city{Sydney}
  \country{Australia}}
\email{ying.zhang@uts.edu.au}

\author{Justin Gooding}
\affiliation{%
  \institution{Australian Centre for Nanomedicine, The University of New South Wales}
  \city{Sydney}
  \country{Australia}}
\email{justin.gooding@unsw.edu.au}

\author{Toby Walsh}
\affiliation{%
  \institution{School of Computer Science and Engineering, The University of New South Wales}
  \city{Sydney}
  \country{Australia}}
\email{t.walsh@unsw.edu.au}







\renewcommand{\shortauthors}{XXXXXX.}

\begin{abstract}
Catalysts are essential for accelerating chemical reactions and enhancing selectivity, which is crucial for the sustainable production of energy, materials, and bioactive compounds. 
Catalyst discovery is fundamental yet challenging in computational chemistry and has garnered significant attention due to the promising performance of advanced Artificial Intelligence (AI) techniques.
The development of Large Language Models (LLMs) notably accelerates progress in the discovery of both homogeneous and heterogeneous catalysts, where their chemical reactions differ significantly in material phases, temperature, dynamics, etc. 
However, there is currently no comprehensive survey that discusses the progress and latest developments in both areas, particularly with the application of LLM techniques. 
To address this gap, this paper presents a thorough and systematic survey of AI-empowered catalyst discovery, employing a unified and general categorization for homogeneous and heterogeneous catalysts. 
We examine the progress of AI-empowered catalyst discovery, highlighting their individual advantages and disadvantages, and discuss the challenges faced in this field.
Furthermore, we suggest potential directions for future research from the perspective of computer science. 
Our goal is to assist researchers in computational chemistry, computer science, and related fields in easily tracking the latest advancements, providing a clear overview and roadmap of this area. 
We also organize and make accessible relevant resources, including article lists and datasets, in an open repository at~\url{https://github.com/LuckyGirl-XU/Awesome-Artificial-Intelligence-Empowered-Catalyst-Discovery}.

\end{abstract}




\received{20 February 2025}

\maketitle

\section{Introduction}
\label{sec1}

\begin{figure} 
    \centering
    \includegraphics[width= 0.9\textwidth]{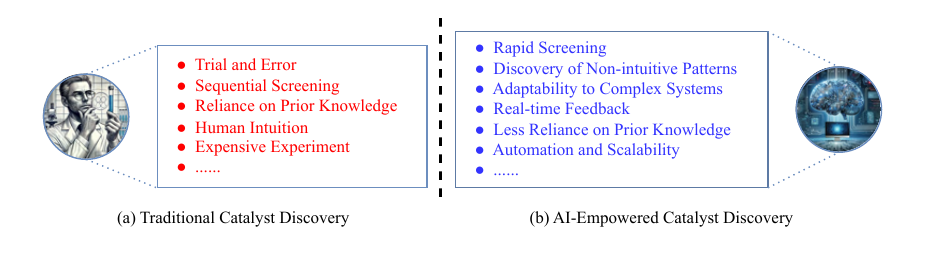}
    \caption{Comparison between traditional catalyst discovery (left) and AI-empowered catalyst discovery (right).}\label{fig:1}
\end{figure}

\subsection{Background}
Catalysts are substances that increase the rate of a chemical reaction without being permanently altered in the process. Catalysis is a highly complex, multiscale phenomenon involving chemical and energy transformations at active sites~\cite{robbins2011simple,senkan1998high,zhong2020accelerated,taylor2015catalysts,hoveyda2007remarkable,zhao2017enhanced,chen2021ph,chen2021key,poerwoprajitno2022single}.
Catalysts play a crucial role in various areas, such as chemical manufacturing~\cite{rase2000handbook,heveling2012heterogeneous,vermeiren2009impact,perez2003formation,gawande2020carbon,gloag2024co}, environment engineering~\cite{wu2015ferric,chen2021heterogeneous,chen2023challenges,chen2022comparative,zhao2018oxidation,yang2021toward,sondergaard2022highly}, and pharmaceutical development~\cite{busacca2011growing,han2021asymmetric,wilcken2013principles,hayler2018pharmaceutical,aleman2013applications,wu2022rational}.
Traditionally, catalyst discovery is predominantly conducted through trial-and-error experiments guided by human intuition and prior knowledge.
However, such trial-and-error methods rely heavily on prior knowledge from experts.
Meanwhile, these methods are time-consuming due to the sequential nature of synthesizing, analyzing, and testing compounds for their desired properties, as shown in Fig.~\ref{fig:1}a.
For instance, early catalyst development involves screening over $2,500$ compositions to identify the optimal catalyst for ammonia synthesis~\cite{potyrailo2011combinatorial}. As computational chemistry advances, high-throughput experiments have been used for the development of general catalyst discovery. Researchers manage to simulate the density functional theory (DFT)~\footnote{Density Functional Theory (DFT) is a computational quantum mechanical modeling technique widely utilized in computational chemistry to examine the electronic structure of many-body systems, including atoms, molecules, and condensed phases. By employing mathematical functions of the electron density—DFT enables the determination of various properties of many-electron systems. 
} by capturing the essential aspects of complex real-world systems, thereby guiding the rational design of experiments~\cite{hannagan2021first,xu2018universal,guo2020tackling,tan2019electrocatalytic,yu2023breaking}.

Recently, there has been a shift towards probing the underlying processes of catalysis to infer design knowledge and develop strategies for creating high-performance catalytic materials using various machine learning (ML) methods~\cite{robbins2011simple,goldsmith2018machine,ozin2018catalyst} due to the following reasons.
First, the number of possible catalyst structures and compositions can be astronomical. ML methods can navigate this vast search space more efficiently than DFT-based methods. Second, catalyst discovery involves dealing with vast amounts of complex, multiscale data, including molecular structures, reaction conditions, and performance metrics. ML methods can efficiently extract meaningful patterns from these data, which would be challenging and time-consuming for human researchers. Third, ML methods can predict potential catalysts based on historical data and known knowledge, which allows researchers to focus their experimental efforts on the most promising candidates and then significantly speed up the discovery process. Therefore, various ML techniques can help discover novel and high-performance catalysts, saving labor and resources, and automating much of the process (see Fig.~\ref{fig:1}b).

Existing modeling paradigms in catalyst discovery evolve from classical learning methods to advanced approaches such as large language models (LLMs). 
Early classical methods~\cite{li2020adaptive,weng2020simple,zhong2020accelerated,sun2020accelerating,mazheika2022artificial,wang2023accelerated,mok2023data,yin2024machine}, such as the regression models, are employed to predict and optimize catalyst performance based on historical data. 
These models are typically trained on datasets containing various catalyst features (e.g., material composition, surface area, porosity) and corresponding performance metrics (e.g., conversion rate, selectivity, stability).
However, classical methods typically have limited performance due to their reliance on feature engineering as well as non-linear interactions among catalyst-related data. To more accurately represent high-order interactions, researchers use graph structures~\cite{gasteiger2021gemnet,wander2022catlas,wang2023dr,yang2023curator,gao2023revisiting} to model the complex interactions between different atoms, where nodes represent atoms and edges represent bonds or atomic neighbors.
With the constructed graphs, graph neural networks (GNNs) are extensively used to model higher-order and complex interactions, such as those between triplets or quadruplets of atoms, which are crucial for accurate predictions. 
However, GNN-based methods require precise atomic coordinates to construct graph representations, while it remains challenging to integrate multiple attributes during modeling, such as space velocity and temperature.



More recent studies employ textual representations to describe adsorbate-catalyst systems as an alternative. 
Textual descriptions offer a natural and human-interpretable way to incorporate various observable features~\cite{sprueill2023monte,lai2023artificial,ock2023catalyst,m2024augmenting,bandeira2024co2}. 
Researchers leverage powerful LLMs to comprehend these textual inputs and predict catalyst properties. 
This innovative approach represents a promising frontier in catalyst discovery, particularly useful given the vast possibilities for catalyst compositions and the complex nature of catalytic reactions.


\begin{table}[t]
    \centering
    \resizebox{\linewidth}{!}{\begin{tabular}{c|c|cc|cccccc}
    \hline
   & &\multicolumn{2}{c|}{Tasks} & \multicolumn{4}{c}{Techniques}\\
   \cline{3-8}
   References& Date& Homogeneous. &Heterogeneous.& Classical M.  &Gene. \& Rein. L. & Graph NNs  & LLMs  \\ 
   \hline
   Freeze et al.~\cite{freeze2019search}&2019&\ding{52}&\ding{52}&\ding{52}& && \\  
    \hline
   Foscato et al.~\cite{foscato2020automated}& 2020&\ding{52}&&\ding{52} & &&\\
    \hline
    McCullough et al.~\cite{mccullough2020high}&2020&&\ding{52}&\ding{52}& \\
    \hline
    Zhu et al.~\cite{zhu2023data}&2023&&\ding{52}&\ding{52}&\ding{52} &&\\
    \hline
    Mou et al.~\cite{mou2023bridging}&2023&&\ding{52}&\ding{52}&&\ding{52}& \\
   \hline
    Ours&2024&\ding{52}&\ding{52}&\ding{52}&\ding{52}&\ding{52}&\ding{52}\\
     \hline
    \end{tabular}}
    \caption{A brief summary of related surveys on catalyst discovery. Note: Homogeneous. and Heterogeneous. denote homogeneous and heterogeneous catalysts, respectively. Classical M., Gene. \& Rein. L., Graph NNs and LLMs denote classical models, generative and reinforcement learning, graph neural networks, and large language models, respectively. }
    \label{tab:1}
\end{table}

\subsection{Motivation}
We present the motivation of this survey paper from two perspectives: the diversity and complexity of existing solutions as well as the limitations of related surveys. 
As introduced previously, existing solutions in catalyst discovery are diverse in terms of the input data types and designed learning methods.
Data inputs are either constructed or reconstructed in different types, such as sequences, (knowledge) graphs, and textual formats. 
According to these data types, researchers explore different learning methods and modeling paradigms for representation learning. 
These methods consequently serve various downstream tasks, such as energy prediction, activity estimation, and stability analysis.
The diversity and complexity of these methods present challenges for researchers in related fields, such as computer science, hindering their ability to effectively learn and apply these techniques.

We summarize related surveys in Table~\ref{tab:1}. 
The survey paper~\cite{foscato2020automated} reviews works on homogeneous catalyst discovery, introducing automated catalyst design techniques and covering methods ranging from classical approaches to various shallow neural networks. 
In contrast, the survey papers~\cite{mccullough2020high,zhu2023data,mou2023bridging} provide a comprehensive comparison of works exclusively on heterogeneous catalyst discovery, discussing techniques and experimental evaluations.
Freeze et al.~\cite{freeze2019search} attempt to offer a unified discussion of discovering both types of catalysts within computational chemistry. 
However, all these surveys focus solely on classical machine learning and neural network models, overlooking advancements in LLM-based methods~\cite{sprueill2023monte,lai2023artificial,ock2023catalyst,m2024augmenting,bandeira2024co2}. 
The emergence of large language models (LLMs) marks a significant milestone in artificial intelligence, becoming increasingly powerful tools across various domains~\cite{birhane2023science,wei2022emergent,jablonka2024leveraging,boiko2023autonomous,white2023future}, including computational chemistry. Given this, it is essential to discuss the progress of LLM-based methods in catalyst discovery. 
Additionally, we offer future directions of AI-empowered catalyst discovery for the first time, which is crucial for improving the efficiency and effectiveness of automated catalyst discovery systems.
Hence, current surveys fail to capture recent strides in catalyst discovery, particularly the integration of LLMs in the design and evaluation of catalysts.

Considering the rapid advancements in catalyst research and the limitations of existing surveys, there is a pressing need for a comprehensive review that covers both homogeneous and heterogeneous catalyst discovery within computational chemistry. 
This paper also offers a comprehensive review of machine learning techniques, ranging from classic models to the latest advancements in large language models.
This review not only summarizes the latest findings but also identifies existing gaps and proposes future directions for catalyst research. 
By addressing these needs, we aim to provide a holistic perspective on the current state of catalyst discovery and emphasize the potential of various advanced approaches and automated systems from both computational chemistry and computer science views in this field.

\subsection{Contribution}
In this paper, we present a comprehensive survey on catalyst discovery from the perspectives of computational chemistry and computer science, covering the discovery of both homogeneous and heterogeneous catalysts. 
We introduce a systematic classification schema to effectively organize the diverse range of existing catalyst discovery models.
Specifically, we categorize the existing research into four main groups based on techniques and modeling: classical, generative and reinforcement learning-based, graph neural network-based, and large language model-based catalyst discovery. 
For each group, we thoroughly review and summarize current discovery methods, highlighting their strengths and limitations. 
We conduct an in-depth analysis of potential research challenges in catalyst discovery and identify promising future research directions for potential advancements. 
Our survey aims to serve as a foundational resource for researchers, helping understand the landscape of computational approaches to catalyst discovery and inspiring further innovations in this critical area of study.

The main contributions of this survey can be summarized as follows:

\begin{itemize}
\item \textbf{Systematic Survey.} We provide a comprehensive review of Artificial Intelligence-empowered catalyst discovery, covering background knowledge, machine learning-based approaches, benchmark datasets, modeling approaches, evaluation methods, and available resources.

\item \textbf{Holistic View.}
We propose a holistic categorization for catalyst discovery, covering both homogeneous and heterogeneous catalysts through direct and inverse designs.
For each category, we discuss their advantages and limitations from technical and chemical views.

\item \textbf{Future Directions.} We summarize and discuss current challenges and limitations and outline potential future research directions in catalyst discovery within the computational chemistry area. Additionally, we suggest potential solutions for several open issues.
\end{itemize}

The roadmap of this survey is as follows: Section~\ref{sec:2} presents the preliminaries of catalyst discovery and categorizes the existing AI-empowered catalyst discovery methods. 
Sections~\ref{sec:3}-~\ref{sec:6} review the primary research problems in each category and analyze their respective strengths and limitations. 
Section~\ref{sec:7} offers an overview of the predominant benchmark datasets, commonly used evaluation metrics, and real-world automated systems. 
Section~\ref{sec:8} discusses the current challenges and proposes five future directions. 
Finally, the survey is concluded in Section~\ref{sec:9}.

\section{Preliminaries and Our Categorisation}\label{sec:2}
Catalysis refers to the acceleration of a chemical reaction by a substance called a catalyst, which remains unchanged after the reaction~\cite{anderson2012catalysis}. Factors such as mixing, surface area, and temperature are crucial for catalytic efficiency. Catalysts typically interact with reactants to form intermediates, producing the final product while regenerating the catalyst. Catalyst discovery spans fundamental scientific research, resource efficiency, and industrial applications, which has the potential to revolutionize chemical processes as well as promote sustainability across multiple domains. 
In this section, we introduce the definitions of catalyst discovery and automated catalyst discovery, discuss data preparation, and present our proposed categorization.

\subsection{Definition of Catalyst Discovery}
Here, we provide specific definitions for homogeneous catalyst discovery and heterogeneous catalyst discovery.

\noindent\textbf{Homogeneous Catalyst Discovery.}
Homogeneous catalyst discovery~\cite{crabtree1999combinatorial,chen2023challenges} focuses on catalysts that are present in the same phase as the reactants, typically in either a gaseous or liquid state. These catalysts are typically dissolved in a solvent along with the substrates, allowing for uniform distribution and interaction at the molecular level. This property facilitates precise control over reaction conditions and can lead to high selectivity and efficiency in chemical processes.


\noindent\textbf{Heterogeneous Catalyst Discovery.}
Heterogeneous catalyst discovery~\cite{hutchings2009heterogeneous} involves catalysts that exist in a different phase than the reactants, typically solid catalysts interacting with liquid or gaseous substrates. These catalysts feature active sites—specific atoms or crystal faces such as edges, surfaces, or steps—where the catalytic activity occurs. These active sites play a critical role in facilitating the chemical reaction by providing a surface for the reactants to adsorb, react, and then desorb as products.


\subsection{Automated Catalyst Discovery}
Catalyst design is a nonlinear optimization problem where changes in catalyst properties, such as activity and selectivity, do not correlate linearly with changes in the catalytic system. This system is defined by numerous parameters, including the catalyst itself, substrates, solvents, and potential additives. The vast number of parameters leads to a combinatorial explosion, making the problem highly complex. Automated catalyst discovery aims to leverage computational tools and high-throughput experimental techniques to systematically and efficiently identify, design, and optimize catalysts.

To manage this complexity, the search space is typically restricted to key elements, often focusing solely on the molecular catalyst and substrates, while excluding solvents and additives. It is crucial to prune these parameters while considering the complexity, computational cost, and performance. 
Ideally, the search space should be dynamically defined, incorporating new knowledge acquired during the design process. 
The catalytic performance $P$ is related to the system parameters by a forward operator $F$, which is usually unknown. 
There are two main strategies in molecular design, that is, \textit{direct design} and \textit{inverse design}~\cite{foscato2020automated,freeze2019search}.

\noindent\textbf{Direct design} in catalyst discovery starts with a defined set of parameters for the catalyst system and uses these parameters to estimate the output performance. 
This method employs an approximate operator $F$ to predict output performance based on the parameters of candidates. 
The process is iterative, akin to traditional experimental approaches where hypotheses are tested. Then, the results guide subsequent experiments. 
The iterative search for optimal catalysts in direct design typically employs heuristic techniques, such as trial and error, to incrementally improve catalyst performance. 
The primary goal is to modify and optimize the parameters until the desired catalytic activity and selectivity are achieved.

\noindent\textbf{Inverse design}, in contrast, starts with the desired optimal performance and works backward to determine the key parameters to achieve this performance. 
This approach effectively reverses the causality relationship defined by the operator $F$. 
Since $F$ is typically not invertible, inverse design involves adding constraints or incorporating performance-driven feedback to make the process feasible. 
Common techniques used in inverse design include high-throughput screening, evolutionary-driven global optimization, and reinforcement learning. 
These methods iteratively evaluate candidate catalysts and adjust their parameters based on performance feedback, aiming to efficiently reach the target performance.


\subsection{Data Preparation}\label{sec:2.3}
\begin{figure}[t]
    \centering
    \includegraphics[width= 0.9\textwidth]{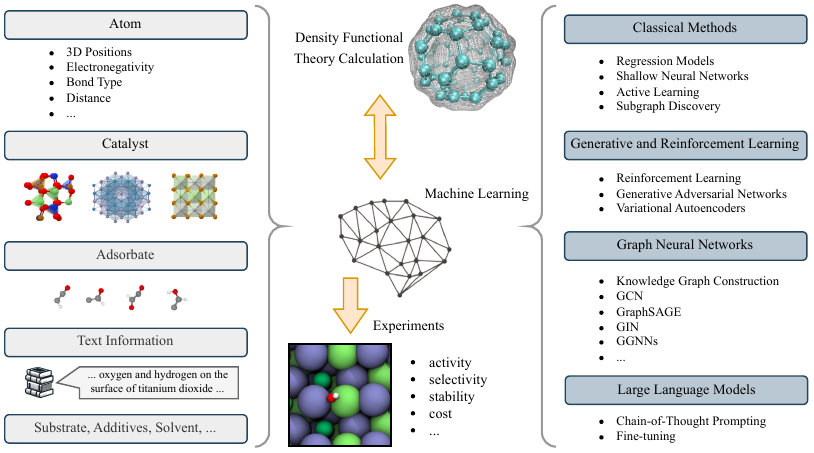}
    \caption{Key Features and Algorithms in Catalyst Design.}
    \label{fig:workflow}
\end{figure}

Data preparation is a critical initial phase in the machine learning pipeline for catalyst discovery. Researchers transform raw chemistry data into powerful features that drive machine learning models. These features are then used as inputs to design and optimize catalysts by leveraging various machine learning techniques, as illustrated in Fig.~\ref{fig:workflow}. This phase involves extracting, creating, and transforming raw data into discriminative and informative features that significantly enhance the performance of machine learning models. The feature engineering for catalyst discovery includes domain knowledge integration~\cite{schlexer2019machine}, extraction of basic features~\cite{castleman2009clusters,ding2019tuning}, creation of derived features~\cite{durand2019computational,mok2023data,zhu2012treespan,margraf2023exploring,gao2023revisiting}, dimensionality reduction~\cite{abdi2010principal,wold1987principal,van2008visualizing}, feature selection~\cite{guyon2003introduction,saeys2007review,xu2018semi,xu2023learning}, transformations and encoding~\cite{ba2016layer,craig2006scaling,church2017word2vec}, and feature engineering for complex data~\cite{gu2018recent,chowdhary2020natural}. For example, extraction of basic features refers to elemental features (e.g., atomic number, atomic mass, electronegativity, and ionization energy) and compositional features (e.g., alloys and compounds). In summary, effective feature engineering enables the identification of key properties that are mostly associated with catalytic activity, stability, and selectivity, thereby facilitating the design of high-performance catalysts for various chemical reactions.

\begin{figure}
    \centering
\begin{adjustbox}{max width=\textwidth}
\begin{forest}
for tree={
    grow=east,
    draw,
    edge= ultra thick,
    parent anchor=east,
    child anchor=west,
    edge path={
        \noexpand\path [draw, \forestoption{edge}] (!u.parent anchor) -- +(10pt,0) |- (.child anchor)\forestoption{edge label};
    },
    anchor=west,
    calign=center,
    l sep+=10pt,
    s sep+=2pt,
    where level=1{tier=word}{},
    font= \scriptsize 
},
where level=0{%
        text width=5cm, 
    }{},
    where level=1{%
        text width=5.5cm, 
    }{},
    where level=2{%
        text width=5cm, 
    }{},
    where level=3{%
        text width=5cm, 
    }{}
[AI-Empowered Catalysts Discovery
[Large Language Model-based Methods, edge=red
[Fine-tuning Based,edge=red
    [Pretained RoBERTa\\ Fine-tuning~\cite{ock2023catalyst},edge=red
    ]]
    [Prompting Based, edge=red,
    [Monte Carlo Tree Search\\ Chain-of-Thought (CoT) Prompting~\cite{sprueill2023monte},edge=red
    ]
    [Chain-of-Thought (CoT) Prompting~\cite{m2024augmenting}, edge=red
    ]
    [Prompting\\ Rule-Based Information Extraction~\cite{bandeira2024co2},edge=red
    ]
    [Active Learning\\ Prompting~\cite{lai2023artificial},edge=red
    ]]
  ]
    [Graph Neural Network-based Methods, edge=blue
    [Complex Graph Representation Based, edge=blue
     [Knowledge Graph Construction\\ Graph Convolution Networks~\cite{gao2023revisiting}, edge=blue]
     [Graph Representation\\ Transformer~\cite{park2023artificial}, edge=blue]
    ]
    [Geometric Graph Neural Network Based, edge=blue
     [Geometric Graph Neural Network\\ Active Learning~\cite{yang2023curator}, edge=blue]
     [Geometric Graph Learning\\ Kohn-Sham Charge-Density~\cite{pope2023towards}, edge=blue]
     [Geometric Graph Neural Network\\ Equivariant Neural Networks~\cite{gasteiger2021gemnet}, edge=blue]
    [Geometric Graph Neural Network\\ Graph Parallelism~\cite{sriram2021towards}, edge=blue]
    [Geometric Graph Neural Network\\ Pretraining Technique~\cite{wander2022catlas}, edge=blue]
    ]
    [Graph Convolution Network Based, edge=blue
     [Graph Isomorphism Neural Network\\ Label Deconstruction and Reconstruction~\cite{wang2023dr}, edge=blue]
    [Graph Convolution Network\\ Voronoi Tessellation~\cite{korovin2023boosting}, edge=blue]
    [Graph Convolution Network\\ Theory-Infused Neural Networks~\cite{pillai2023interpretable}, edge=blue]
    [General Graph Neural Network\\ Physical Method~\cite{duval2024phast}, edge=blue]
    ]
    ]
    [Generative and Reinforcement Learning-based Methods, edge=green
        [Reinforcement Learning, edge=green
            [Deep Q-Network\\ Objective Sampling~\cite{lacombe2023adsorbrl}, edge=green]
            [Deep Reinforcement Learning\\ Markov Decision Process~\cite{dos2021navigating}, edge=green]
        ]
        [Recurrent Neural Network\\ Variational Autoencoder~\cite{schilter2023designing}, edge=green]
        [Generative Adversarial Network~\cite{ishikawa2022heterogeneous}, edge=green]
    ]   
    [Classical-based Methods, edge=purple
    [Active Learning~\cite{mok2023data, zhong2020accelerated,yin2024machine, li2020adaptive}, edge=purple]
        [Regression Models, edge=purple
            [Decision-Tree Regression\\ Subgraph Discovery~\cite{mazheika2022artificial}, edge=purple]
            [Symbolic Regression~\cite{weng2020simple}, edge=purple]
            [Gaussian Process Regression\\ Active Learning~\cite{yin2024machine, li2020adaptive}, edge=purple]
            [Tree Regression~\cite{wang2023accelerated,sun2020accelerating}, edge=purple]
        ]
        [Shallow Neural Networks~\cite{jiang2023automated}, edge=purple]
    ]
]
\end{forest}
\end{adjustbox}
    \caption{Holistic categorisation of representative works in AI-empowered approaches for catalyst discovery.}
    \label{fig:2}
\end{figure}
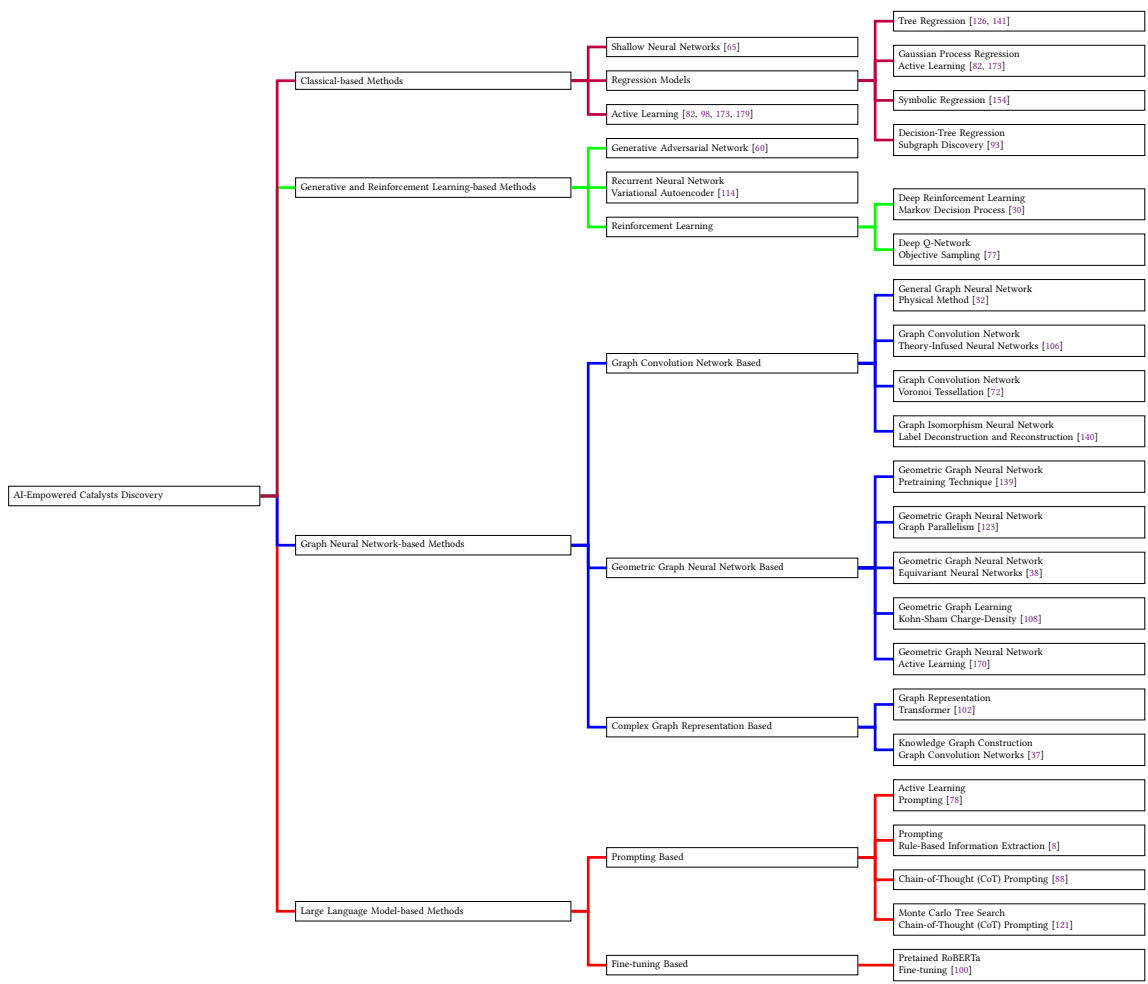

\subsection{Our Proposed Categorization}
In this paper, we provide a comprehensive review and discussion of the existing works in catalyst discovery within computational chemistry and computer science, focusing particularly on the application of various learning techniques. Our goal is to integrate and unify research efforts across both homogeneous and heterogeneous catalyst discovery, encompassing approaches in both direct design and inverse design. 
As shown in Fig.~\ref{fig:2}, we categorize the existing works into four distinct groups based on the learning techniques: classical methods, generative and reinforcement learning-based methods, graph neural network-based methods, and large language model-based methods.

In the following sections, we delve into each category, examining the specific research task, detailed techniques, and their relative advantages and limitations. 
Additionally, we provide a thorough evaluation of these methods, highlight key developments, and identify future research directions in catalyst discovery.

\section{Classical Methods in Catalysts}\label{sec:3}
\subsection{Overview}
In this section, studies~\cite{petras2002age,maity2021multivariate} demonstrate the transformative potential of combining classical machine learning with chemistry for catalyst discovery. By integrating computational predictions with experimental validation, these studies present robust methodologies for identifying and optimizing catalysts with superior performance. Specifically, a classical model (e.g., Gaussian process regression) is trained on a dataset containing input features (e.g., elemental properties, structural descriptors) and corresponding output values (e.g., catalytic activity, adsorption energies), as introduced in Section~\ref{sec:2.3}. These classical models are then used to predict properties such as adsorption energies, reaction rates, and catalytic activities from input features derived from the composition and structure of materials. This approach enables researchers to efficiently explore large chemical spaces, identify key factors influencing catalytic performance, and accelerate the development of high-performance catalysts. Next, we present the widely used classical methods in catalyst discovery, providing detailed techniques, their advantages and limitations.

\subsection{Technique Details of Classical Methods}

\subsubsection{Gaussian Process Regression.} Gaussian Process Regression (GPR)~\cite{li2020adaptive,zhong2020accelerated} is used to model the relationship between input features (e.g., composition, electronic structure) and output properties (e.g., adsorption energies). It captures uncertainty in predictions, guiding the selection of the most promising perovskite candidates for further investigation. Given training data $\{x_i, y_i\}$, the predictive distribution for a new point $x^*$ is
\begin{equation}
p(f(x^*) \mid x^*, x, y) = \mathcal{N}(\mu^*, \sigma^{*2}),
\end{equation}
where $\mu^* = k(x^*, x) \left[K(x, x) + \sigma_n^2 I\right]^{-1} y$ and $\sigma^{*2} = k(x^*, x^*) - k(x^*, x) \left[K(x, x) + \sigma_n^2 I\right]^{-1} k(x, x^*)$. $k(x^*, x)$ is the kernel function, $K(x, x)$ is the covariance matrix, and $\sigma_n^2$ is the observational noise. Last, we can achieve the function $f(\cdot)$ using the GPR model. In addition, GPR is used with active learning to iteratively select the most informative data points for DFT calculations, optimizing the discovery process of Cu-Al electrocatalysts~\cite{zhong2020accelerated}. It helps predict adsorption energies and identify optimal catalytic sites.

\subsubsection{Symbolic Regression.} Symbolic regression aims to describe the relationship between input features (e.g., number of $d$ electrons, electronegativity) and the target property (e.g., oxygen evolution reaction (OER) activity)~\cite{weng2020simple,mazheika2022artificial,wang2023accelerated}. This approach provides interpretable models that can reveal underlying physical relationships governing catalytic activity. We formulate it as 
\begin{equation}
y = f(x_1, x_2, \ldots, x_n) + \epsilon,
\end{equation}
where $f(\cdot)$ is a symbolic expression and $\epsilon$ represents the observational noise. $f(\cdot)$ can derive from either compositional data or elemental properties according to different functions. 

\begin{figure}[t] 
    \centering
    \includegraphics[width= 0.99\textwidth]{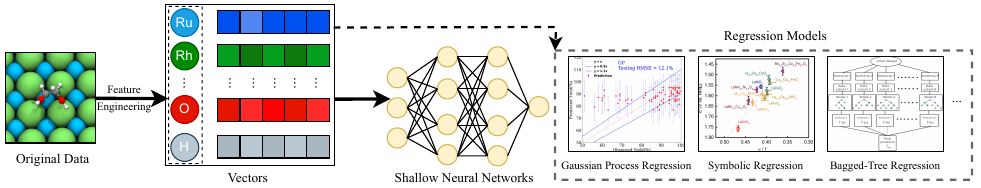}
    \caption{Overview of classical methods in catalyst discovery. Regression models are collected from~\cite{weng2020simple,kovavcevic2021construction,mishra2023predicting}.}
    \label{fig:rl_nn}
\end{figure} 
\subsubsection{Bagged-Tree Regression.} Bagged-tree regression aggregates predictions from multiple decision trees to enhance accuracy and generalization ability. It is used to predict adsorption energies and electronic structures of graphdyine (GDY)-based atomic catalysts, providing reliable insights into their hydrogen
evolution reaction (HER) performance~\cite{sun2020accelerating}. The model can be formulated as
\begin{equation}
\hat{y} = \frac{1}{B} \sum_{b=1}^{B} \hat{y}^{(b)},
\end{equation}
where $\hat{y}^{(b)}$ is the output prediction from the $b$-th tree. 
Bagged-tree regression models build multiple decision trees and average their predictions to reduce the model's variance, thereby improving prediction accuracy.


\subsubsection{Shallow Neural Networks.} Neural networks learn complex, non-linear relationships between input features (e.g., elemental compositions, DFT data) and output properties (e.g., OER overpotential)~\cite{jiang2023automated}.
This aids in the automated synthesis and optimization of OER catalysts using Martian meteorites. A typical shallow neural network consists of an input layer, one hidden layer, and an output layer. The input layer receives the data, the hidden layer processes the data linearly and then applies a non-linear activation function, and the output layer produces the final prediction (see Fig.~\ref{fig:rl_nn}). We formulate the process as
\begin{equation}\label{eq:3.6}
    \mathbf{z}_{1} = \mathbf{W}_{1} \mathbf{x} + \mathbf{b}_{1},\;\mathbf{a}_{1} = \sigma(\mathbf{z}_{1}),\;\mathbf{z}_{2} = \mathbf{W}_{2} \mathbf{a}_{1} + \mathbf{b}_{2},\;\mathbf{y} = \sigma(\mathbf{z}_{2}), 
\end{equation}
where $\mathbf{W}_{1}$ and $\mathbf{W}_{2}$ are the weight matrices for the hidden and output layers, respectively. $\mathbf{b}_{1}$ and $\mathbf{b}_{2}$ are the bias vectors for the hidden and output layers, respectively. $\sigma(\cdot)$ represents the activation function. Despite their simplicity, these networks are powerful tools in machine learning, capable of handling various problems effectively in catalyst discovery.


\subsubsection{Active Learning}
Active learning is a training way to improve the machine learning interatomic potentials (MLIPs). Existing methods~\cite{yang2023curator,yin2024machine,mok2023data,mazheika2022artificial,li2020adaptive} leverages active learning for the construction of strong model potentials (e.g., regression model). Among these works, we focus on batch active learning, which can efficiently identify the most informative structures from production simulations and expand their applicability across a border chemical space. With active learning, the model can iteratively select batches of unlabeled data to be labeled by an oracle (usually a human annotator) and then incorporated into the training set. For the batch section methods, there is a benchmark over the MD17 dataset, which presents their performance of different methods~\cite{larsen2017atomic}.

    
    




\begin{table}[t]
    \centering
    \resizebox{\linewidth}{!}{\begin{tabular}{c|c|m{3.5cm}|m{4.5cm}|m{4.5cm}}
    \hline
        Ref. & Year & Techniques & Merits & Demerits  \\
        \hline
         \citep{li2020adaptive} & 2020& 1. Gaussian process regression
         
        2.  Active learning
        &1. Identify known and unknown candidates
        
        2. Provide molecular orbital insights 
        & 1. High prediction uncertainty 
        
        2. Significant computational costs\\
        \hline
         \citep{weng2020simple} & 2020& 1. Symbolic regression 

         2. Genetic programming
        &1. Provide interpretable mathematical formulas
        
        2. Accelerate the screening and discovery process
        & 1. Suffer from formability and stability issues
        
        2. Dependent on the quality of training data \\
         \hline
         \citep {zhong2020accelerated} & 2020& 1. Active learning

         2. Gaussian process regression
        &1. Exhibit enhanced selectivity
        
        2. Systematically discover and optimize multi-metallic catalysts 
        & 1. Dependency on the initial dataset
        
        2. Poor generalization ability\\
         \hline
         \citep{sun2020accelerating} & 2020& 1. Bag tree algorithm
         
         2. Fuzzy models
        &1. Combination of advantages of both DFT and ML
        & 1. High computation cost 
        
        2. Limited prediction ability \\
        \hline
        \citep{mazheika2022artificial} & 2022& 1. Decision-tree rsegression
        
        2. Subgraph discovery
        &1. Provide interpretable formulas 
        
        2. Handle large chemical space  
        & 1. Require high-throughput DFT calculations 
        
        2. Unstable prediction outputs\\
         \hline
         \citep{wang2023accelerated} & 2023& 1. Tree regression 
         
        &1. Inductive learning
        
        2.  Relationship between elemental properties and catalytic performance
        & 1. Suffer from interpretability issues
        
        2. Insufficient or biased data \\
         \hline
         \citep{jiang2023automated} & 2023& 1.Shallow neural networks
        2. Bayesian optimization 
        &1. Handle large chemical spaces 
        
        2. Good adaptability for different data distribution
        & 1. Require substantial computational resources for MD simulations
        
        2. Environmental constraints\\
         \hline 
         \citep {mok2023data} & 2023& 1. Active motifs-based machine learning 
        2. XGBoost regression 
        &1. Consider both activity and selectivity
        
        2. Explore a vast chemical space
        & 1. Hard to handle high-dimensional data
        
        2. Predictive limitations\\
         \hline
         \citep{yin2024machine} & 2024& 1. Gaussian process regression
         
        2. Active learning
        &1. Minimize the number of required experiments
        
        2. Capable of screening large datasets
        & 1. Poor generalization ability
        
        2. Limited interpretability\\
         \hline
    \end{tabular}}
    \caption{Comparative study of classical catalyst discovery.}
    \label{tab:2}
\end{table}

\subsubsection{Subgraph Discovery}
Subgraph discovery (SGD)~\cite{atzmueller2015subgroup,kuramochi2001frequent,wang2024neural,WangVLDB,luo2021,qin2015locally,wang2022reinforcement,wang2022neural} is typically used to identify patterns within a dataset by finding subgraphs that exhibit significantly different characteristics compared to the overall population. 
These subgraphs are defined by specific conditions or rules, called selectors, that describe the values of certain features within the dataset. 

The catalysts can be naturally modeled as molecular graphs where the nodes are the atoms and edges are the bonds between atoms.
In catalyst discovery, subgraph discovery aims to find combinations of material properties (referred to as "material genes") that correlate with desired catalytic performance indicators, such as activation energies or reaction rates.
Here, we take subgraph discovery in~\cite{mazheika2022artificial} as an example and introduce its usage. 
The predicted activation energies and reaction rates are used to discover which combinations of material features are associated with the activation of CO2 on semiconductor oxide surfaces. 
The quality function $F(Z)$ is used to evaluate the quality of a subgraph $Z$ based on its target property (e.g., OCO-angle or C–O bond length). 
We formulate the minimization of OCO-anglea as
\begin{equation}
F(Z) = \theta_{\text{cut}} \frac{s(Z)}{s(Y)} \left( \frac{\max(Z) - \alpha_g}{\min(Y) - \alpha_g} \right) \cdot u(p),
\end{equation}
where $\theta_{\text{cut}}$ is the Heaviside step function that ensures all data points in the subgroup meet the cutoff condition. 
$s(Z)$ and $s(Y)$ are sizes of the subgroup $Z$ and entire dataset $Y$, respectively. 
$\max(Z)$ and $\min(Y)$ are the maximum value of the target property in $Z$ and minimum value in $Y$, respectively. $\alpha_g$ is the gas-phase value of the target property, and $u(p)$ is the factor accounting for the Sabatier principle, increasing the quality of subgroups with appropriate adsorption energies. Then we formulate the C–O bond length maximization as
\begin{equation}
F(Z) = \theta_{\text{cut}} \frac{s(Z)}{s(Y)} \left( \frac{\min(Z) - l_g}{\max(Y) - l_g} \right) \cdot u(p),
\end{equation}
where $l_g$ is the gas-phase value of the C–O bond length. For the selectors, they are conditions that define a subgroup. They are defined as inequalities involving features $f_i$:
\begin{equation}
\text{Selector} = (f_1 < a) \text{ AND } (f_2 \geq b) \text{ AND } \ldots,
\end{equation}
where $f_i$ are features of the material and $a, b$ are threshold values determined by the SGD algorithm. With the subgroup discovery algorithm, we first identify primary features that are related to the target property and then apply SGD to find subgraphs with optimal target property values. Last, we analyze the identified subgroups to understand the material features that correlate with the desired catalytic performance.


\subsection{Summary}
As shown in Table~\ref{tab:2}, the comparative analysis of various classical machine learning techniques for catalyst discovery reveals a landscape of promising methodologies, each with distinct strengths and weaknesses. These techniques offer significant benefits such as enhanced selectivity, interpretability, and the ability to manage large chemical spaces, which are crucial for accelerating the discovery and optimization of high-performance catalysts. For example, regression techniques such as Gaussian process regression~\cite{li2020adaptive} and tree regression~\cite{sun2020accelerating} offer the ability to predict properties from input features and provide interpretable formulas. However, they also reveal some disadvantages, including dependence on the quality and quantity of input data, the computational intensity of advanced models, and the necessity for extensive experimental validation to confirm predictions. Despite these limitations, integrating computational and experimental approaches continues to demonstrate transformative potential.

\begin{tcolorbox}[colback=blue!5!white, colframe=blue!75!black, left=1.5mm, 
    right=1mm, 
    top=1mm, 
    bottom=1mm, 
    boxrule=1mm, 
    leftrule=1mm, 
    rightrule=0mm, 
    toprule=0mm, 
    bottomrule=0mm, 
    arc=0mm 
]
\textbf{Takeaway 1:} Classical methods, grounded in mathematical frameworks, offer strong interpretability, making them valuable in searching for useful catalyst materials from large chemical space. However, these approaches are typically tailored to narrow, well-defined tasks, limiting their scalability to complex, high-dimensional problems. Future research should focus on designing lightweight models that integrate mathematical methodologies to strike an optimal balance between DFT and ML in this line. By leveraging physics-informed ML and hybrid modeling techniques, researchers can enhance both computational efficiency and interpretability
\end{tcolorbox}

\section{Generative and reinforcement learning in Catalysts}\label{sec:4}
\subsection{Overview}
\begin{figure} 
    \centering
    \includegraphics[width= 0.99\linewidth]{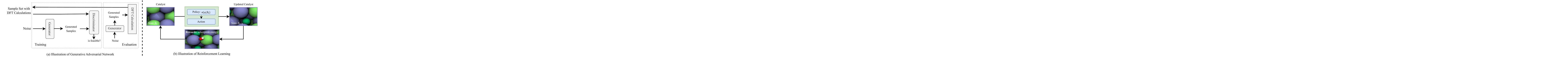}
    \caption{Overview of generative and reinforcement learning in catalyst discovery \citep{lacombe2023adsorbrl}.}
    \label{fig:RL}
\end{figure}

Generative and reinforcement learning-based catalyst discovery integrates machine learning algorithms (e.g., deep Q-network~\cite{kaelbling1996reinforcement,arulkumaran2017deep} and Generative Adversarial Networks (GANs)~\cite{wang2017generative,goodfellow2020generative}) with computational chemistry to design and optimize new catalytic materials. Concretely, reinforcement learning leverages the iterative decision-making process, where an agent learns to optimize catalyst performance by receiving feedback from the environment, while generative learning typically generates novel catalyst configurations through a competitive process between a generator and a discriminator (see Fig.~\ref{fig:RL}). 
These approaches effectively process comprehensive datasets on catalyst compositions, structures, and performance metrics, leading to the creation and refinement of catalysts with enhanced properties not present in the initial dataset. 
By generating new data with properties similar to the training data, these methods accelerate the discovery of high-performance catalysts.
Next, we elaborate on the technical details for generative and reinforcement learning in catalyst discovery, focusing on three key techniques: generative adversarial networks, variational autoencoders, and reinforcement learning.

\subsection{Technique Details of Generative and Reinforcement Learning}

\subsubsection{Generative Adversarial Networks}
Generative Adversarial Networks (GANs) are a class of machine learning frameworks designed for generating new data samples that are indistinguishable from real (ground truth) data~\cite{wang2017generative,goodfellow2020generative}. As introduced by~\cite{goodfellow2020generative}, GANs consist of two neural networks, i.e., the generator and the discriminator, which are trained simultaneously through adversarial processes. The generator aims to generate data, while the discriminator attempts to distinguish between ground truth and generated data. This competitive dynamic drives both networks to improve their performance until the generated data becomes nearly identical to the ground truth data. The generator network takes random noise as input and produces data samples. The discriminator network, on the other hand, evaluates these samples along with ground truth data, providing feedback to the generator. Through backpropagation, the generator learns to produce increasingly realistic data by minimizing the discriminator's ability to distinguish between ground truth and fake samples. This adversarial training loop continues until the generator produces high-quality data that closely simulates the distribution of the real dataset.

Mathematically, the GAN framework is formulated as a minimax game between the generator $G$ and the discriminator $D$. The objective function $V(G, D)$ is defined as:
\begin{equation}\label{eq:4.18}
\min_G \max_D V(G, D) = \mathbb{E}_{x \sim p_{\text{data}}(x)} [\log D(x)] + \mathbb{E}_{z \sim p_z(z)} [\log (1 - D(G(z)))],
\end{equation}
where $p_{\text{data}}(x)$ is the distribution of the ground truth data, and $p_z(z)$ is the distribution of the noise input to the generator. The generator $G(\cdot)$ aims to minimize this objective, while the discriminator $D(\cdot)$ aims to maximize it.

Based on Eq.~\eqref{eq:4.18}, GANs offer a revolutionary approach to designing new catalytic materials with enhanced properties. Traditional methods of catalyst discovery~\cite{senkan1999discovery,senkan1998high} often involve extensive experimental trial-and-error processes, which can be time-consuming and resource-intensive. GANs, however, can automate and expedite this process by generating novel catalyst configurations that are likely to exhibit high performance. Below, we take the~\cite{ishikawa2022heterogeneous} as an example and present the learning process of GANs in catalyst discovery:

\noindent\textbf{Data Preparation}: A dataset comprising initial metal surfaces with known compositions and DFT-calculated turnover frequencies (TOFs) for a target catalytic reaction, such as NH$_3$ formation, is prepared. The dataset includes various atomic configurations and their corresponding performance metrics.

\noindent\textbf{GAN Training}: The GAN is trained using the data, where the generator network proposes new catalyst surfaces, and the discriminator network evaluates their feasibility based on the TOF values. The training process aims to generate surfaces with higher TOFs that are not present in the initial data.

\noindent\textbf{Iterative Improvement}: The GAN-DFT process involves continuously generating new catalyst surfaces using the generator, evaluating them with DFT calculations, and adding high-performing surfaces to the data. This iterative loop allows the GAN to refine its proposals, progressively improving the catalytic performance of the generated surfaces.

\noindent\textbf{Validation and Optimization}: The generated catalysts are validated through further DFT calculations to confirm their performance. The goal is to discover new catalytic materials that exhibit significantly higher TOFs, enhancing the efficiency of the target catalytic reaction.

For each generated catalyst surface $G(z)$, the TOF is calculated using DFT-based microkinetic analysis. The TOF calculation considers several elementary reactions with reaction energies determined by DFT:
\begin{equation}
\text{TOF} = \frac{k \cdot p_{\text{N}_2} \theta_{\text{vac}}}{1 + \sum_i \theta_i},
\end{equation}
where $k$ is the rate constant of the rate-determining step, $p_{\text{N}_2}$ is the partial pressure of nitrogen, $\theta_{\text{vac}}$ is the fractional coverage of vacant sites, and $\theta_i$ are the coverages of adsorbed species. The generator's goal is to maximize the TOF by generating optimal catalyst surfaces. The iterative improvement process can be defined as
\begin{equation}
\max_G \; \text{TOF}(G(z)) \quad \text{subject to} \quad D(G(z)) \approx 1,
\end{equation}
where $D(G(z)) \approx 1$ indicates that the discriminator classifies the generated surface as realistic. GANs in catalyst discovery, not only accelerate the identification of high-performance catalysts but also open up new possibilities for designing materials with tailored properties for specific reactions. 


\subsubsection{Variational Autoencoders.}
Variational Autoencoders (VAEs)~\cite{kingma2019introduction,doersch2016tutorial} are a class of generative models widely used for various applications, including the design of catalysts. 
A VAE consists of two main components: an encoder and a decoder. 
The encoder compresses input data, e.g., tokenized representation of potential catalysts (i.e., SMILES~\cite{weininger1988smiles} or SELFIES~\cite{krenn2020self}) into a lower-dimensional latent space; 
meanwhile, the decoder reconstructs the original data from this latent representation. 
With these latent representations, we can leverage the property predictor (e.g., feed-forward neural network) to accomplish the downstream tasks, such as reaction energy prediction. 
This architecture allows for efficient sampling and generation of new data points, which is crucial in designing new catalysts with desired properties.
Next, we introduce the encoder and decoder functions in catalyst discovery~\cite{schilter2023designing}.

\noindent\textbf{Encoder.} The encoder $q_{\phi}(z|x)$ maps the input data $x$ (e.g., molecular representations) to a latent space $z$. This is often parameterized using neural networks:
\begin{equation}\label{eq:5.2.3.1}
z \sim q_{\phi}(z|x) = \mathcal{N}(\mu_{\phi}(x), \sigma_{\phi}(x)^2),
\end{equation}
where $\mu_{\phi}(x)$ and $\sigma_{\phi}(x)$ are the mean and standard deviation predicted by the encoder. We formulate them as
\begin{equation}
    \mu_{\phi}(x) = \operatorname{NN}_{\mu}(x),\; \sigma_{\phi}(x) = \operatorname{NN}_{\sigma}(x),
\end{equation}
where $\operatorname{NN}_{\mu}(\cdot)$ and $\operatorname{NN}_{\sigma}(\cdot)$ can be any neural networks. 
Based on Eq.~\eqref{eq:5.2.3.1}, we can encode the catalysts data (e.g., SMILES string) to the latent space.

\noindent\textbf{Decoder.} The decoder $p_{\theta}(x|z)$ reconstructs the input data from the latent space $x' \sim p_{\theta}(x|z)$. The decoder is also parameterized by neural networks, aiming to produce data that is similar to the original input. We formulate it as
\begin{equation}
p_{\theta}(x|z) = \operatorname{NN}_{\theta}(z),
\end{equation}
where $\operatorname{NN}_{\theta}(\cdot)$ can be any neural networks. 
For example, in~\cite{schilter2023designing}, Oliver et al. leverages recurrent neural networks to instantiate the $\operatorname{NN}(\cdot)$. 
The decoder function is to decode the latent representation to reconstruct the original data such as the SMILES string.

\noindent \textbf{Learning Objective.} The VAE is trained to minimize the evidence lower bound (ELBO), which consists of two terms: the reconstruction loss and the Kullback-Leibler (KL) divergence:
\begin{equation}\label{eq:5.2.3.5}
L(\theta, \phi; x) = \mathbb{E}_{q_{\phi}(z|x)} [\log p_{\theta}(x|z)] - \beta \, D_{KL} (q_{\phi}(z|x) \| p(z)),
\end{equation}
where $p(z)$ is a standard normal distribution $\mathcal{N}(0, I)$. Based on the learning objective in Eq.~\eqref{eq:5.2.3.5}, we can predict binding energies from the latent space and optimize the latent space to generate new catalysts with desired properties.

\begin{table}[t]
    \centering
    \resizebox{\linewidth}{!}{\begin{tabular}{c|c|m{3.5cm}|m{4.5cm}|m{4.5cm}}
    \hline
        Ref. & Year & Techniques & Merits & Demerits  \\
        \hline
        \citep{dos2021navigating} & 2021& 1. Reinforcement learning &1. Comprehensive analysis and model-building
        
        2.  Automated design, make, test, analyze cycle&1. Strong data dependency\\
        \hline
        \citep{ishikawa2022heterogeneous} & 2022&1. Generative
adversarial network&1. Discover new and high-performance catalysts

2. Automatic proposal and optimization of new catalytic materials &1. Both DFT calculations and GAN training are computationally intensive

2. Neglect surface stability\\
        \hline
         \citep{lacombe2023adsorbrl} & 2023& 1. Deep Q-Network
         
        2.  Objective sub-sampling
        &1. Capable of optimizing adsorption energies for multiple target adsorbates simultaneously
        
        2. Improve performance in environments with sparse rewards
        & 1. Suffer from the local minima, especially in multi-objective setups\\
        \hline
        \citep{schilter2023designing} &2023&1. Recurrent neural network
         
         2. Variational autoencoder &1. Achieve high validity in generating new catalyst structures &1. Neglect reaction conditions (e.g., solvents, temperature)\\
         \hline
    \end{tabular}}
    \caption{Comparative study of generative and reinforcement learning-based catalyst discovery.}
    \label{tab:3}
\end{table}


\subsubsection{Reinforcement Learning}
Reinforcement Learning (RL) is to learn to make decisions by interacting with an environment given the optimization condition~\cite{kaelbling1996reinforcement,arulkumaran2017deep}. The primary objective of reinforcement learning is to determine the optimal policy for an agent to take actions based on its current state, such that it maximizes the cumulative reward over time. This process is typically formulated using the framework of Markov Decision Processes (MDPs)~\cite{van2012reinforcement}, which provide a mathematical formalization of decision-making situations where outcomes are partly random and partly under the control of a decision-maker. Generally, RL includes five main components: States (S), Actions (A), Rewards (R), Policy ($\pi$), and Value Functions. A crucial concept in RL is the Bellman Equation~\cite{gottipati2020learning,zhou2017optimizing}, which provides a recursive decomposition for solving the optimal policy and value functions. The Bellman equation for the value function $V^\pi(s)$ is defined as
\begin{equation}
     V^\pi(s) = \sum_{a \in A} \pi(a|s) \sum_{s', r} p(s', r | s, a) \left[ r + \gamma V^\pi(s') \right],
\end{equation}
where $V^\pi(s)$ represents the value of being in state $s$ under policy $\pi$, $\pi(a|s)$ is the probability of taking action $a$ in state $s$ under policy $\pi$, $p(s', r | s, a)$ is the probability of transitioning to state $s'$ and receiving reward $r$ after taking action $a$ in state $s$, and $\gamma$ is the discount factor which balances the importance of immediate and future rewards.

Similarly, the Bellman equation for the action-value function $Q^\pi(s, a)$ is given by
\begin{equation}
    Q^\pi(s, a) = \sum_{s', r} p(s', r | s, a) \left[ r + \gamma \sum_{a'} \pi(a'|s') Q^\pi(s', a') \right],
\end{equation}
where $Q^\pi(s, a)$ represents the value of taking action $a$ in state $s$ under policy $\pi$.

These equations are fundamental in developing algorithms for learning optimal policies. Common RL algorithms include Q-learning~\cite{watkins1992q} and policy gradient methods~\cite{sutton1999policy}, each of which employs these principles to iteratively improve the policy.

In the context of catalyst discovery, we take the~\cite{lacombe2023adsorbrl,dos2021navigating} as an example and introduce how to design the RL model to find the new catalysts:

\noindent\textbf{States ($S$)}: Representations of catalysts, such as unary or ternary compounds from a dataset like the \textit{Materials Project}. Each state is a combination of atomic elements forming the catalyst, represented by a low-dimensional one-hot vector corresponding to the elements.

\noindent\textbf{Actions ($A$)}: Steps the agent can take to modify or traverse the dataset of materials, such as removing elements.

\noindent\textbf{Rewards ($R$)}: Feedback based on the catalytic performance, such as adsorption energy ($E_{\text{ads}}$) for a target adsorbate. Rewards are formulated to encourage the agent to find optimal catalysts, e.g., $r = -E_{\text{ads}}$ to find strong binding states.

\noindent\textbf{Policy ($\pi$)}: The strategy used by the agent to select actions based on the current state. Policies are trained using algorithms like Q-learning~\cite{lacombe2023adsorbrl}. Here, the Bellman equation for the Q-value is defined as
\begin{equation}
Q^*(a|S) = r(a|S) + \gamma \max_a \left( Q^*(a|S') \right).
\end{equation}

\noindent\textbf{Multi-objective Goal-Conditioning}: For more complex scenarios, policies are trained to meet multiple objectives simultaneously. An objective vector $g$ encodes the goals for different adsorbates, and the reward function is adjusted accordingly:
\begin{equation}
\pi(a|S) = \pi(a|S, g),\; r(S, g) = f\left( E_{\text{ads}}(S_{t+1}, \text{ads}_i) , g \right)
\end{equation}

The goal-conditioned Bellman equation becomes:
\begin{equation}
Q^*(a|S, g) = r(a|S, g) + \gamma \max_a \left( Q^*(a|S', g) \right).
\end{equation}

\noindent\textbf{Evaluation Metrics}: The effectiveness of the trained policies is evaluated by measuring the average adsorption energy of the final states. The improvement $\Delta$ from initial to final states is calculated as
\begin{equation}
\Delta = \frac{1}{N} \sum_{i \in \text{final}} \left( -E_{\text{ads}}(S_i) \right) - \frac{1}{N} \sum_{j \in \text{initial}} \left( -E_{\text{ads}}(S_j) \right).
\end{equation}

Based on the above process, RL models are employed to predict and optimize the adsorption energies of catalysts. The reward functions are designed to steer the agent towards the most promising materials, while policy training ensures that the agent makes decisions at each state. 
Then the effectiveness of the trained policies is evaluated by various metrics, such as the improvement in average adsorption energy. By automating the search for high-performance catalysts, RL reduces the dependency on traditional experimental methods, accelerating the discovery of catalysts. 

 \noindent\textbf{Remarks.} Generative and reinforcement learning have been extensively utilized in computational chemistry. These approaches not only expedite the discovery of new compounds but also significantly cut down the reliance on traditional, resource-intensive experimental methods. For a deeper understanding and additional insights into other computational chemistry tasks, please refer to related surveys~\cite{studebaker1957chemistry,sridharan2024deep,vanhaelen2020advent,kell2020deep,chuang2020learning}.

\subsection{Summary}
In summary, generative and reinforcement learning can significantly enhance catalyst discovery by optimizing catalytic properties through efficient exploration and generating novel catalyst configurations beyond existing datasets. We provide a comparison of these approaches in Table~\ref{tab:3}. RL excels in efficiently navigating complex chemical spaces and handling multi-objective optimization, which reduces the need for exhaustive searches and allows for continuous improvement through iterative learning. However, RL requires significant computational resources, faces challenges with sparse rewards, and involves complex implementation. GANs and VAE, on the other hand, can extrapolate beyond initial datasets to generate new catalyst materials, automating the design and evaluation process and continuously refining outputs through iterative training. Nonetheless, generative learning-based methods also demand substantial computational costs, and their effectiveness is heavily dependent on the quality and diversity of the initial dataset. Additionally, GANs can experience training instability and their models can be difficult to interpret, complicating the understanding of the underlying mechanisms. By integrating these techniques with computational chemistry, researchers can push the boundaries of traditional experimental methods and open new avenues for innovation in catalysis research. 
\begin{tcolorbox}[colback=blue!5!white, colframe=blue!75!black, left=1.5mm, 
    right=1mm, 
    top=1mm, 
    bottom=1mm, 
    boxrule=1mm, 
    leftrule=1mm, 
    rightrule=0mm, 
    toprule=0mm, 
    bottomrule=0mm, 
    arc=0mm 
]
\textbf{Takeaway 2:} Reinforcement learning can accelerate the search of chemical spaces for catalyst discovery, particularly in multi-objective optimization. Generative learning, on the other hand, enables the design of novel catalyst materials by leveraging existing knowledge. The efficiency of both generative and reinforcement learning methods can be significantly enhanced when guided by expert knowledge, as it helps improve data quality, refine optimization objectives, and reduce computational overhead.
\end{tcolorbox}

\section{Graph Neural Networks in Catalysts}\label{sec:5}
\subsection{Overview}
\begin{figure} 
    \centering
    \includegraphics[width= 0.9\textwidth]{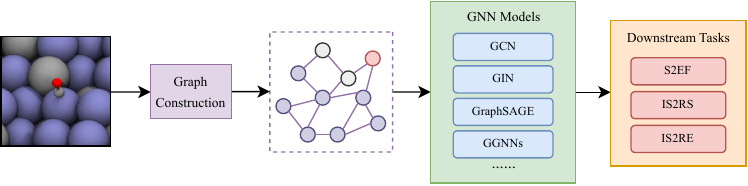}
    \caption{Overview of graph neural networks in catalyst discovery.}
    \label{fig:gnn}
\end{figure}

Graph Neural Networks (GNNs) are powerful in analyzing graph structure data in learning complex correlations with wide applications in various areas, such as biology~\cite{mao2023predicting,li2022inferring,long2022pre,jha2022prediction} and computational chemistry~\cite{duvenaud2015convolutional,kearnes2016molecular,wang2022molecular,hao2020asgn}.
Along this way, researchers attempt to represent catalytic systems as graphs, where atoms are depicted as nodes and interactions between them as edges. 
These graphs effectively capture the complex interdependencies within atomic structures, allowing for accurate prediction of catalytic properties. 
The procedure typically involves three main steps in Fig.~\ref{fig:gnn}: graph construction, embedding learning, and performance prediction. In the graph construction phase, a molecule's atomic structure is converted into a graph format, integrating geometric and directional information to enhance the representation. During the learning phase, various GNNs~\cite{wu2020comprehensive,wang2021gognn,xu2018powerful,wu2023billion,he2020lightgcn,xu2024timesgn,xu2024scalable} can be trained on extensive datasets, such as the Open Catalyst 2020 (OC20), to learn complex correlations and then generate node and edge embeddings. The model is trained to minimize the difference between its predictions and known values from the training data and then predict properties such as energy and forces. 
Last, the trained GNN predicts the properties of new catalytic systems, providing valuable insights into their performance and guiding the design of more efficient and effective catalysts. Next, we elaborate on the technical details of graph neural networks in catalyst discovery.

\subsection{Technique Details of Graph Neural Networks}
Here, chemical data are first transferred into graph structures. Then, we leverage graph neural networks to generate node and edge representations for different downstream tasks, such as activity prediction~\cite{pillai2023interpretable}. 
Here, we first introduce the definition of graphs and then give a brief summary of the existing GNN-related techniques.

\subsubsection{Definition of Graph} 
\begin{definition}[Graph.]\label{def:4.1.1}
A graph is represented as $\mathcal{G}= \{\mathcal{V}, \mathcal{E}\}$, where $\mathcal{V}$ is the set of nodes and $\mathcal{E}$ is the set of edges. Let $v_i\in \mathcal{V}$ be a node and $e_{ij} = (v_i, v_j)\in \mathcal{E}$ be an edge pointing from $v_j$ to $v_i$. The neighborhood of a node $v$ is denoted as $\mathcal{N}(v) = \{u\in\mathcal{V}|(v, u)\in \mathcal{E}\}$. 
\end{definition}

Based on Definition~\ref{def:4.1.1}, we introduce knowledge graph construction from catalyst data.




\subsubsection{Knowledge Graph Construction from Catalyst Data} In the context of catalyst discovery, researchers often model catalyst data into knowledge graphs that can represent atomic attributes or reaction semantics, thus facilitating the systematic analysis of catalyst data. Constructing a knowledge graph from catalyst data involves several key steps. 
We take the literature~\cite{gao2023revisiting} as an example, in which researchers collect Cu-based electrocatalyst data and transform these data into the knowledge graph for subsequent GNN modeling.

\noindent (1) Data Collection and Preprocessing: Researchers begin by compiling a comprehensive dataset of scientific literature on Cu-based electrocatalysts for CO2 reduction. 
This dataset includes titles, abstracts, and full texts of relevant papers. Then, they extract and clean the textual data to ensure consistency and remove irrelevant information. This process involves parsing the text to identify key sections such as materials, regulation methods, products, Faradaic efficiency, cell setup, electrolyte, synthesis methods, current density, and voltage.

\noindent (2) Named Entity Recognition (NER): Researchers use a SciBERT-based framework~\cite{beltagy2019scibert} to perform NER on the processed text. This model is specifically trained to recognize entities such as materials, regulation methods, products, and Faradaic efficiency. They implement a BiLSTM-CRF model~\cite{chen2017improving} to tag and classify these entities within the text. SciBERT provides contextual embeddings for each word, which are then processed by the BiLSTM to capture the sequential context. The CRF layer assigns the final entity labels, ensuring accurate identification and classification of relevant terms.

\noindent (3) Relationship Extraction: Researchers identify relationships between the recognized entities by analyzing how different entities, such as specific materials and their methods, are connected based on their co-occurrence and contextual proximity in the text. They typically use similarity metrics to quantify the strength of these relationships, allowing them to understand the interactions and dependencies between various elements in the dataset.

\noindent (4) Graph Construction: Last, researchers construct the knowledge graph where nodes represent the entities (e.g., materials, methods, and products) and edges represent the relationships between these entities. Meanwhile, the graph is stored in a graph database, which allows for efficient querying and visualization of the relationships.

By following these steps, the knowledge graph provides a structured and comprehensive way to represent and analyze the vast amount of information available in the scientific literature on Cu-based electrocatalysts for CO2 reduction. This approach enables researchers to gain deeper insights based on the knowledge graphs and various advanced knowledge graph techniques~\cite{wang2017knowledge,ji2021survey,lin2015learning,wu2024query2gmm}, which help make better decisions in catalyst design and optimization.

\subsubsection{Graph Neural Network (GNN)} Generally speaking, graph neural networks (GNNs) are formulated using two main functions: the aggregation function and the update function. Following this paradigm, there are typical cases of GNN, including Graph Convolution Neural Network (GCN)~\cite{kipf2022semi}, GraphSAGE~\cite{hamilton2017inductive}, and Graph Isomorphism Network (GIN)~\cite{xu2018powerful}. We formulate the paradigm as 
\begin{equation}
    \text{Aggregation:}\quad \bf{n}_{v}^{(l)} = \operatorname{AGGREGATE}_{l}(\{\bm{h}_u^{(l)}, \forall u\in \mathcal{N}(v)\}),\; \text{Update:}\quad \bf{h}_{v}^{(l+1)} = \sigma(\bm{W}^{(l)}[\bm{h}_v^{(l)}\oplus\bf{n}_{v}^{(l)}]),
\end{equation}
where $\operatorname{AGGREGATE}_{l}(\cdot)$ denotes the aggregation function at the $l$-th layer and $\sigma(\cdot)$ is the non-linear activation function and $\bm{W}^{(l)}$ is the learnable transformation matrix for layer $l$. For the graph neural networks, the aggregation step sums up the features of the neighboring nodes $\mathcal{N}(v)$ for each node $v$. The update step combines the original node feature $\mathbf{h}_v^{(l)}$ with the aggregated neighbor features $\mathbf{n}_{v}^{(l)}$. The combined features are then transformed using a learnable matrix $\mathbf{W}^{(l)}$ and passed through a non-linear activation function $\sigma(\cdot)$ to update node representation $\mathbf{h}_{v}^{(l+1)}$. Looking into the catalyst discovery, these graph neural networks are typically used to predict the energy of catalytic systems within the Open Catalyst Project dataset, in which node features can represent atomic properties (e.g., electronegativity, period, and group position) and edge features can represent contact solid angles and types. These methods can significantly improve the final error that equals $651$ meV on the Open Catalyst Project dataset~\cite{korovin2023boosting}.

\subsubsection{Geometric Graph Neural Networks (GGNNs)}
GGNNs~\cite{schutt2021equivariant,gasteiger2021gemnet,gasteiger2019directional,pope2023towards} is designed to predict tensorial properties and molecular spectra by leveraging equivariant message passing. Unlike traditional graph neural networks that rely on rotationally invariant representations, GGNNs employ rotationally equivariant representations to better capture directional information and interactions between atoms in three-dimensional space. The core of GGNNs involves the message passing between nodes (atoms) in a graph (molecule). The message-passing for 3D-embedded graphs can be represented as
\begin{equation}
    m_{i}^{(t+1)} = \sum_{j \in N(i)} M(s_i^{(t)}, s_j^{(t)}, \vec{r}_{ij}), \;
    s_i^{(t+1)} = U(s_i^{(t)}, m_{i}^{(t+1)}),
\end{equation}
where $s_i^{(t)}$ is the state of node $i$ at time step $t$, $\vec{r}_{ij}$ is the relative position vector between nodes $i$ and $j$, $M(\cdot)$ is the message function, and $U(\cdot)$ is the update function. For rotationally equivariant message passing, the message and update functions are modified to fit vectorial features:
\begin{equation}\label{eq:6.2.5.3}
  \vec{m}_{i}^{(t+1)} = \sum_{j \in N(i)} \vec{M}(s_i^{(t)}, s_j^{(t)}, \vec{v}_i^{(t)}, \vec{v}_j^{(t)}, \vec{r}_{ij}),\;
    s_i^{(t+1)} = U(s_i^{(t)}, m_{i}^{(t+1)}),\;
    \vec{v}_i^{(t+1)} = \vec{U}(s_i^{(t)}, \vec{m}_{i}^{(t+1)}).  
\end{equation}

Here, $\vec{v}_i^{(t)}$ represents the vectorial features of node $i$ at time step $t$. Based on Eq.~\eqref{eq:6.2.5.3}, GGNNs can extend traditional message passing by incorporating vectorial features, allowing the model to handle directional information.

\begin{table}[!ht]
    \centering
    \resizebox{0.95\textwidth}{!}{\begin{tabular}{c|c|m{4cm}|m{4.5cm}|m{4.5cm}}
    \hline
        Ref. & Year & Techniques & Merits & Demerits  \\
        \hline
        \citep{gasteiger2021gemnet}& 2021 & 1. Geometric message passing neural network
        
        2. Equivariant neural networks &1. Respect the physical symmetries of the systems 
        
        2. Suitable for a wide range of molecular systems & 1. Neglect the inductive learning \\
         \hline
         \citep{sriram2021towards}& 2021& 1. Directional graph neural networks

2. Graph parallelism
        &1. Large-scale parameters
        
        2. Atomic structure learning 
        & 1. Require an additional backward pass
        
        2. Non-trivial carbon emissions for model training\\
         \hline
        \citep{wander2022catlas} & 2022 &1. Geometric message passing neural network 
        
        2. Pretrained model&1. Improve the reliability of screening results
        & 1. Rely on the quality of the pre-trained GNN models
        
        2. Sensitive to the selection of input parameters and filtering criteria.\\
         \hline 
        \citep{wang2023dr} & 2023& 1. Graph neural networks 
        
    2. Label deconstruction and reconstruction
        & 1. Enhanced supervision signal

        2. Robustness to graph structural variation
        & 1. High computation cost due to deconstruction and reconstruction
        
        2. Bottleneck in equilibrium state prediction problems for atomic systems\\
         \hline
         \citep{gao2023revisiting}&2023 & 1. Knowledge graph construction for Cu-based electrocatalysts
         
         2. Graph convolution networks &1. Provide a comprehensive view of the relationships between catalysts and catalytic activities
         
         2. Provide domain-specific insights&1. Poor cross-domain generalization ability
         
         2. Fail to process large-scale electrocatalytic CO2 reduction task\\
         \hline
         \citep{yang2023curator} &2023 &1. Polarizable atom interaction neural network

    2. Active learning &1. Handle larger systems and longer simulation time

    2. Autonomous workflow that simplifies the process for users &1. Require large datasets and numerous iterations

    2. Dependency on representative molecular structures\\
    \hline
         \citep{korovin2023boosting} & 2023&1. Graph convolution network
         
         2. Voronoi Tessellation &1. Suitable for high-throughput catalyst discovery
         
         2.  Incorporates physically meaningful geometric and chemical information &1. Depend heavily on the quality of the features, such as Voronoi volumes and solid angles
         
         2. Poor generalization capability\\
         \hline
         \citep{pillai2023interpretable} &2023 &1. Graph convolution network
         
         2. Theory-infused neural networks &1. The theory-infused approach provides insights into catalyst performance &1. Generalization limitations in exploring new catalysts \\
         \hline
         \citep{park2023artificial}& 2023& 1. Transformer 

        2. Graph representation
        &1. Provide a versatile way to document and utilize diverse experimental data
        & 1. Depend heavily on the data completeness\\
         \hline
         \citep{pope2023towards}&2024 &1. Geometric graph learning
         
         2. Kohn-Sham Charge-Density &1. Generalize to new combinations of elements not seen during training
         
         2. Fast convergence& 1. Higher inference costs due to the need for detailed charge density predictions\\
         \hline
         \citep{duval2024phast}&2024 &1. Graph convolution network

         2. Physical method
         &1. Fast model training on standard CPU hardware 

         2. Graph construction for catalyst-adsorbate modeling &1. Physics-based properties increase the complexity of the model inference processes\\
         \hline
        \end{tabular}}
    \caption{Comparative study of graph neural network-based catalyst discovery.}
    \label{tab:4}
\end{table}

\subsubsection{Graph Transformer}\label{sec:4.2.4}
The Graph Transformer model~\cite{yun2019graph} is designed to process and analyze graph-structured data by incorporating the principles of the Transformer model~\cite{vaswani2017attention}, which has achieved significant success in natural language processing. 
By leveraging the self-attention mechanism, graph Transformers can capture complex relationships and dependencies in graph data more effectively than traditional graph neural networks. We formulate it as
\begin{equation} \label{eq:5.24}
    \hat{h_t} = \operatorname{MSA}(Q, K, V),\;
    h_t = \operatorname{FFN}(\hat{h_t}\parallel h_t^-),
\end{equation}
where $\operatorname{MSA}(\cdot)$ is the multi-head-attention~\cite{vaswani2017attention} and $\operatorname{FFN}(\cdot)$ is the feed-forward function. Based on Eq.~\ref{eq:5.24}, we generate graph representation for polymeric structures and continuous-flow reactors, which can then be applied to develop highly effective models for designing ROP catalysts and architecturally valid co-polymers~\cite{park2023artificial,korovin2023boosting}.


\subsection{Summary}
Graph Neural Networks (GNNs) have emerged as a transformative tool for catalyst discovery, leveraging their ability to model complex atomic interactions and predict catalytic properties with high accuracy. 
The primary advantage of GNNs lies in their capacity to handle large datasets and incorporate intricate geometric and directional information, enabling accurate predictions of molecular properties such as energy, forces, and reactivity. 
However, the complexity of GNNs also presents challenges, including high computational requirements for model training and inference, as well as dependency on high-quality, extensive training datasets to achieve optimal performance. 
We provide a comprehensive comparison for this category in Table~\ref{tab:4}. 
In summary, GNNs make significant applications in catalyst discovery including the optimization of electrocatalysts for CO2 reduction, ammonia oxidation, and the direct conversion of syngas to valuable chemicals, due to their special ability to explore complex correlations.

\begin{tcolorbox}[colback=blue!5!white, colframe=blue!75!black, left=1.5mm, 
    right=1mm, 
    top=1mm, 
    bottom=1mm, 
    boxrule=1mm, 
    leftrule=1mm, 
    rightrule=0mm, 
    toprule=0mm, 
    bottomrule=0mm, 
    arc=0mm 
]
\textbf{Takeaway 3:} In catalyst discovery, graph-based representations offer an effective means of modeling complex geometric and directional information. Here, nodes encode atomic properties, while edges preserve contact solid angles and interaction types. By leveraging GNNs, researchers can predict the energetic properties of catalytic systems or generate novel catalyst materials by learning intricate atomic and molecular interactions. However, the performance of these methods is highly sensitive to graph construction quality and dataset consistency. Furthermore, generalizing these models across diverse chemical domains remains a significant challenge, as variations in graph representations and reaction conditions can hinder their transferability. In the future, researchers can enhance the generalization capabilities of graph-based catalyst discovery models by incorporating domain adaptation strategies, neighbor encoding techniques, and other advanced methodologies.
\end{tcolorbox}


\section{Large Language Models in Catalysts}\label{sec:6}
\subsection{Overview}
Large Language Models (LLMs) are the latest powerful techniques in artificial intelligence, which have been applied to catalyst discovery. These models are utilized for critical tasks such as extracting relevant chemical information from scientific literature, predicting catalyst properties like adsorption energy, understanding reaction mechanisms, and suggesting optimal parameters. 
LLMs can operate in two primary modes: prompt-based and fine-tuning based modes in catalyst discovery. In the prompt-based approach, LLMs generate responses based on specific queries without additional training, making it quick and useful for exploratory analysis. The fine-tuning based approach involves further training the LLMs on domain-specific datasets, enhancing their accuracy for specific tasks related to catalyst discovery. The learning process includes pre-training on a general text corpus, followed by fine-tuning with specialized chemical data for the fine-tuning approach or developing effective prompts for the prompt-based approach. LLM-based methods can predict potential catalysts and optimize synthesis routes to specific property predictions and reaction mechanism explanations. Here, we introduce the two types of approaches: Prompting and Fine-tuning.

\subsection{Technique Details of Large Language Models}
First, we provide a comprehensive overview, followed by a detailed introduction to these two groups.

\noindent\textbf{Prompt-based Learning.} This approach involves crafting prompts that guide the model to generate the desired output without modifying the underlying model weights. It relies heavily on the design of the prompt and the model's ability to generalize from its pre-trained knowledge to new scenarios in catalyst discovery.

\noindent\textbf{Fine-tuning.} In contrast, fine-tuning involves continuing the training process of a pre-trained model on a specific dataset to adapt its weights to particular tasks in catalyst discovery. This method can yield more accurate and specialized performance on tasks closely related to fine-tuning data.

\subsubsection{Prompting Technique.} Chain-of-Thought (CoT) prompting~\cite{zhang2022automatic} is a technique used to enhance the reasoning capabilities of large language models by guiding them through a structured reasoning process. This method is particularly useful in catalyst discovery~\cite{sprueill2023monte,m2024augmenting}, where complex chemical reasoning and multi-step processes are involved. 
Next, we present a specific example of CoT Prompting in catalyst discovery, which aims to identify an optimal catalyst for CO2 reduction to methanol~\cite{m2024augmenting}.

The problem is first broken down into four steps: 
(1) Identify key properties required for an effective catalyst in CO2 reduction; 
(2) Determine which materials possess these properties; 
(3) Evaluate the stability and reactivity of these materials under reaction conditions; and 
(4) Recommend the most promising catalyst based on the evaluation.
Here, the following intermediate prompts are created:



 


\begin{tcolorbox}[colback=gray!10!white, colframe=gray!70!black, title=Intermediate Prompts]
\begin{itemize}
    \item Prompt 1: What are the key properties that an effective catalyst for CO2 reduction to methanol should have?
    \item Prompt 2: Which materials are known to possess these key properties?
    \item Prompt 3: How stable and reactive are these materials under the conditions of CO2 reduction to methanol? 
    \item Prompt 4: Based on stability and reactivity, which material would you recommend as the most promising catalyst for CO2 reduction to methanol?
\end{itemize}
\end{tcolorbox}

\begin{tcolorbox}[colback=gray!10!white, colframe=blue!90!black, title=Generate Responses for Each Prompt]
\begin{itemize}
    \item Response 1: Key properties include high adsorption energy for CO2, high selectivity for methanol, and stability under reaction conditions.
    \item Response 2: Materials such as copper, zinc, and various metal oxides are known to possess high adsorption energy for CO2 and good selectivity for methanol.
    \item Response 3: Copper-zinc alloys are particularly stable and reactive under CO2 reduction conditions, showing high efficiency in methanol production.
    \item Response 4: Based on the evaluation, a copper-zinc alloy is recommended as the most promising catalyst for CO2 reduction to methanol due to its high stability and reactivity.
\end{itemize}
\end{tcolorbox}

Based on these responses, the optimal catalyst for CO2 reduction to methanol is likely a copper-zinc alloy, as it possesses the necessary properties of high adsorption energy for CO2, high selectivity for methanol, and stability under reaction conditions. 
Last, the recommendation of the LLM is validated against experimental data, after which prompts and responses are further refined to ensure accuracy and reliability.

By following this CoT prompting procedure, LLMs can effectively navigate complex scientific tasks and provide valuable insights for catalyst discovery. Note that prompted-based LLMs provide responses based on the existing knowledge base, thus they cannot generate new theoretical rules but can provide new insights for the given prompts. This structured approach ensures that each aspect of the problem is thoroughly considered, leading to more accurate and actionable outcomes. 

\begin{table}[t]
    \centering
    \resizebox{\linewidth}{!}{\begin{tabular}{c|c|m{3.5cm}|m{4.5cm}|m{4.5cm}}
    \hline
        Ref. & Year & Techniques & Merits & Demerits  \\
        \hline
        \citep{sprueill2023monte} & 2023& 1. Large language models

        2. Monte Carlo tree search
        &1. A more thorough exploration of possible solutions

        2. Zero-shot capabilities
        & 1. LLMs can produce incorrect or ungrounded answers
        
        2. Dependence on reward function\\
         \hline
         \citep{lai2023artificial}&2023&1. Large language model 
         
         2. Active learning 
         
         3. Prompting &1. Large search space &1. The initial sample size and distribution can impact the effectiveness of the optimization process\\
         \hline
         \citep{ock2023catalyst}&2023&1. Large language model

         2. Fine-tuning &1. Use textual inputs that are easily interpretable by humans
         
         2. Demonstrate significant error cancellation& 1. Limited by textual representations
         
         2. Initial performance on unseen adsorbate molecules is lower\\
         \hline
         \citep{m2024augmenting}&2024&1. Large language models
         
         2. Chain-of-Thought prompting &1. Zero-shot capabilities & 1. Require a large number of API calls and significant computational resources\\
         \hline
         \citep{bandeira2024co2}&2024&1. Rule-based information extraction
         
         2. Large language models
         
         3. Prompting &1. Comprehensive data extraction
         
         2. Enable tracking of material usage trends &1. A risk of extracting incorrect or misleading information
         
         2. Rule-based methods can be limited by the specificity of regular expressions\\
         \hline
        \end{tabular}}
    \caption{Comparative study of large language model in catalyst discovery.}
    \label{tab:6}
\end{table}

\subsubsection{Fine-tuning Technique.}

\begin{figure}[t]
    \centering
    \includegraphics[width= 0.6\textwidth]{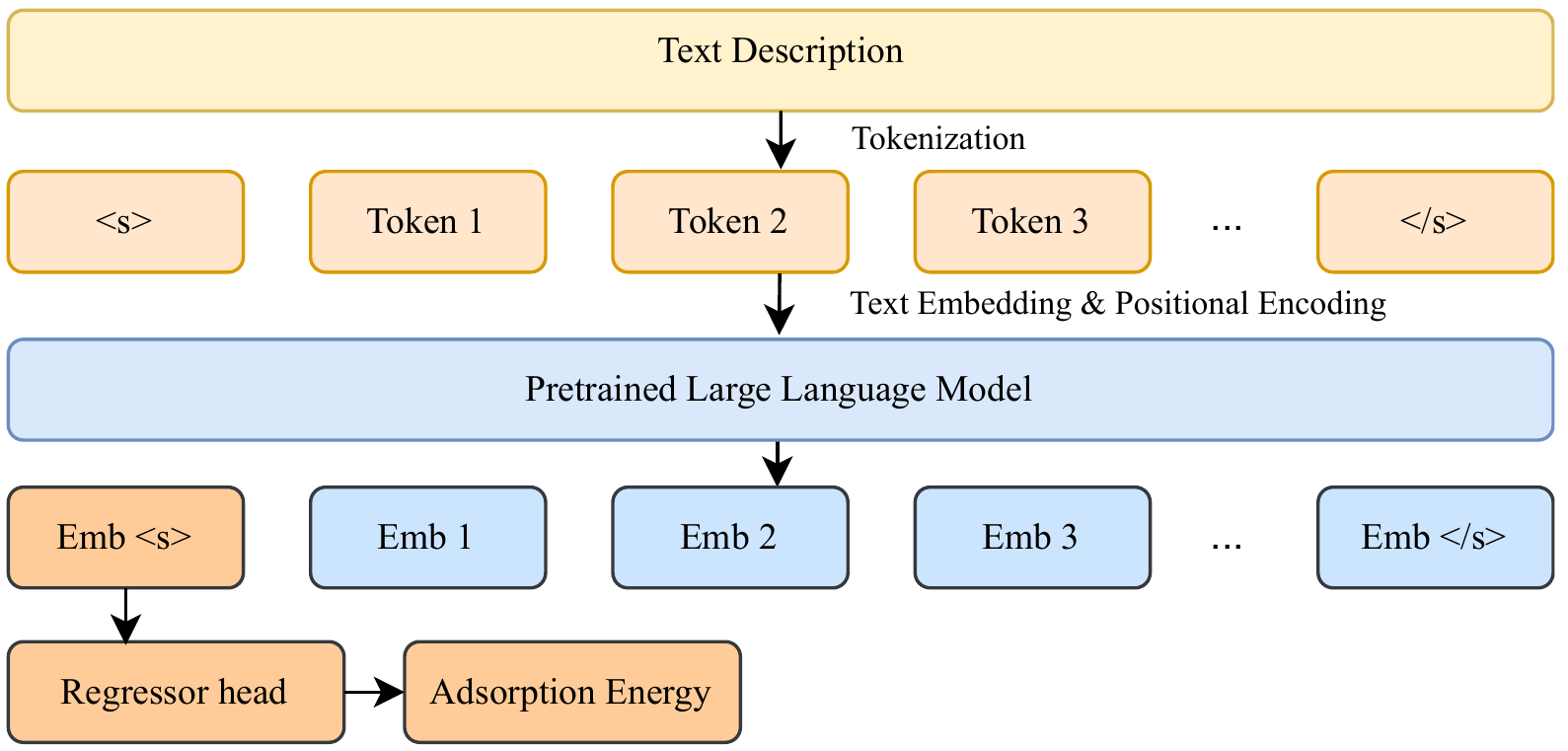}
    \caption{Overview of fine-tuning technique for LLMs in catalyst discovery, in which the term ``Emb" means embeddings.}
    \label{fig:finetuning}
\end{figure} 

Fine-tuning~\cite{bakker2022fine} based models for catalyst discovery involve customizing a pre-trained model with specific data relevant to chemical catalysis, as shown in Fig.~\ref{fig:finetuning}. 
This process enhances the model’s ability to generate accurate and domain-specific predictions, including the following procedures:
\begin{itemize}
    \item Choosing a pre-trained model such as GPT-3~\cite{floridi2020gpt} or LLaMa-2~\cite{touvron2023llama}, which are initially trained on vast amounts of general text data.

    \item Collecting datasets relevant to catalyst discovery, including research articles, experimental data, and chemical databases. 
    There are some specific datasets, such as the Open Catalyst 2020 (OC20) dataset, PubChem, ZINC, and other chemical databases. 
    We need to clean the data to remove noise, irrelevant information, and duplicates, and preprocess the text to ensure it is in a suitable format for training.

    \item With the clean data, we load the pre-trained model and initialize it for fine-tuning with the following steps: (1) Feed the domain-specific data into the model; (2) Set up the training configuration, including the learning rate, batch size, and number of epochs; and (3) Define the loss function (e.g., Mean Squared Error) and optimizer (e.g., AdamW).

    \item Next step is training and evaluation. We train the model on the training data and evaluate the fine-tuned model using metrics such as Mean Absolute Error (MAE). After convergence, we test the model on the test dataset to ensure it generalizes well to unseen data.
\end{itemize}

Fine-tuning adapts a pre-trained LLM to a specific domain, enhancing its ability to make accurate and relevant predictions in that field. 


\subsubsection{Summary}
Large Language Models (LLMs) have emerged as the latest tools in the field of catalyst discovery, offering significant advancements in applications such as extracting chemical information from vast scientific literature, predicting catalyst properties, understanding reaction mechanisms, and optimizing synthesis parameters. The primary advantages of LLMs lie in their ability to process and interpret complex chemical data effectively and accelerate the catalyst discovery process through prompt-based and fine-tuning approaches. However, they also present challenges such as dependency on high-quality, domain-specific datasets, computational resource demands, and potential limitations in accuracy for unseen data. We provide a comparison of existing methods in Table~\ref{tab:6} with its specific advantages and disadvantages. In summary, LLMs will play a crucial role in driving scientific and industrial advancements in chemical research and catalyst optimization, as they can integrate interdisciplinary knowledge and improve data-driven insights.

\begin{tcolorbox}[colback=blue!5!white, colframe=blue!75!black, left=1.5mm, 
    right=1mm, 
    top=1mm, 
    bottom=1mm, 
    boxrule=1mm, 
    leftrule=1mm, 
    rightrule=0mm, 
    toprule=0mm, 
    bottomrule=0mm, 
    arc=0mm 
]
\textbf{Takeaway 4:} LLM-based approaches offer valuable insights for complex scientific challenges that traditional neural models struggle to address, as they can leverage a vast and diverse knowledge base. To maximize their effectiveness in catalyst discovery, these models should be integrated into a human-in-the-loop framework, where collaboration with domain experts enhances accuracy, interpretability, and practical utility. Future research can focus on constructing comprehensive cross-domain datasets that encompass a wide range of chemical systems.
\end{tcolorbox}

\begin{table}[t]
    \centering
    \resizebox{\linewidth}{!}{\begin{tabular}{c|c|m{3.6cm}}
    \hline
    Name & Task & Domain \\
    \hline
    Open Catalyst 2020 & S2EF, IS2RS, IS2RE & Renewable energy and environmental applications \\ 
    \hline
    Open Catalyst 2022 & S2EF-Total, IS2RS, IS2RE-Total & Oxide materials, coverages, and adsorbates \\ 
    \hline
    Catalysis Hub & Energy Prediction & - \\
    \hline
    High Entropy Alloy & Energy prediction & Oxygen reduction reaction \\ 
    \hline
    CO2 Electroreduction & Energy prediction & $\text{CO}_2$ reduction reaction \\
    \hline
    NH3 Formation & Heterogeneous catalyst generation & $\text{NH}_3$ formation reaction\\
    \hline
    Suzuki–Miyaura C–C cross-coupling & Energy prediction & Suzuki–Miyaura C–C cross-coupling reaction \\
    \hline
    Cu-based single-atom-alloyed CO adsorption & Energy prediction & $\text{CO}_2$ reduction reaction\\
    \hline
    GASpy & Energy Prediction & $\text{CO}_2$ reduction and $\text{H}_2$ evolution \\
    \hline
    Uncertainty Benchmark & Energy prediction & - \\
    \hline
    BioFuels Question Reasoning & Prompting optimization & Questions and answers about catalytic reactions\\
    \hline
    \end{tabular}}
    \caption{Dataset description in catalysts.}
    \label{tab:7}
\end{table}

\section{DATASETS, EVALUATION, AND OPEN RESOURCES}\label{sec:7}
In this section, we introduce the commonly used datasets and evaluation metrics for various catalyst discovery tasks and summarize real-world open-source systems in catalyst discovery. 
These resources can support researchers in finding suitable datasets and evaluation metrics to test methods and provide an overview of the practical applications in catalyst discovery.
\subsection{Datasets}\label{sec:9.1}
We collect $11$ datasets specific to different materials and tasks, which are introduced below.
\noindent\textbf{Open Catalyst 2020.} 
The Open Catalyst 2020 (OC20) dataset~\cite{chanussot2021open} features over $1.2 $million DFT relaxations of molecular adsorptions onto surfaces, amounting to approximately $250$ million single-point calculations. 
It includes $82$ distinct adsorbates, such as small adsorbates, $\text{C}_1$/$\text{C}_2$ compounds, and nitrogen/oxygen-containing intermediates, all relevant to renewable energy and environmental applications. These relaxations are conducted on randomly selected low-Miller-index facets of stable materials from the Materials Project, resulting in surfaces derived from $55$ different elements and their combinations.

In the OC20 dataset, there are approximate $872,000$ computed adsorption energies, which include unary catalysts ($3.7\%$), binary catalysts ($61.4\%$), and ternary catalysts ($34.9\%$). Regarding elemental composition, these catalysts may feature reactive nonmetal elements ($28.9\%$), alkali metals ($8.1\%$), alkaline earth metals ($10.2\%$), metalloids ($26.4\%$), transition metals ($81.3\%$), and post-transition metals ($37.2\%$). In terms of adsorbates, $6.6\%$ of the calculations involve adsorbates containing only oxygen or hydrogen, $25.2\%$ involve $\text{C}_1$ adsorbates, $44.4\%$ involve $\text{C}_2$ adsorbates, and $29.0\%$ involve nitrogen-containing adsorbates.

\noindent\textbf{Open Catalyst 2022.} The Open Catalyst 2022 (OC22) dataset~\cite{tran2023open} comprises $62,331$ DFT relaxations, totaling approximately $9,854,504$ single-point calculations, conducted across a variety of oxide materials, coverages, and adsorbates.
It is designed to complement the OC20 dataset that lacks oxide materials, facilitating the development of more generalizable machine learning models for catalysis. 
This dataset encompasses the configurational complexity of oxide surfaces, including various surface terminations, adsorbate\&slab configurations, coverage levels, and non-stoichiometric substitutions and vacancies.

\noindent\textbf{Catalysis Hub.} The Catalysis Hub~\cite{winther2019catalysis} offers an open-access database for chemical reactions on catalytic surfaces, aggregating results from over $50$ publications and datasets.
This extensive database includes more than $100,000$ chemisorption and reaction energies derived from electronic structure calculations. 
It encompasses approximately $700$ different chemical reactions, involving over $100$ adsorbed species and $3,000$ distinct catalytic material surfaces.

\noindent\textbf{High Entropy Alloy.} The High Entropy Alloy dataset~\cite{batchelor2019high} focuses on the catalytic activity of high-entropy alloys in the oxygen reduction reaction. To accurately simulate the randomness of a high-entropy alloy, DFT calculations are performed on the binding of *OH and *O across $871$ and $998$ different $2 \times 2$ unit cells, respectively.

\noindent\textbf{CO2 Electroreduction.} The CO2 Electroreduction dataset~\cite{ma2015machine} is for the $\text{CO}_2$ reduction reaction, which includes small amounts of adsorption energies for $\text{CO}_2$ Electroreduction Catalyst Screening


\noindent\textbf{NH3 Formation.} The NH3 Formation dataset~\cite{ishikawa2022heterogeneous} is for the $\text{NH}_3$ formation reaction over Rh-Ru alloy surfaces, including six elementary reactions: $\text{N}_2$ dissociation, $\text{H}_2$ dissociation, $\text{NH}_\text{x}$ formation (where x ranges from 1 to 3), and $\text{NH}_3$ desorption. In this dataset, DFT calculations are performed to evaluate the reaction energies of these processes.


\noindent\textbf{Suzuki–Miyaura C–C cross-coupling.} The Suzuki–Miyaura C–C cross-coupling dataset~\cite{meyer2018machine} comprises a total of $7,054$ reaction energy values corresponding to descriptors, which is generated by combining $91$ ligands, including CO, phosphine, N-heterocyclic carbene, and pyridine, with six transition metals (Ni, Pd, Pt, Cu, Ag, and Au).

\noindent\textbf{Cu-based single-atom-alloyed CO adsorption.} The Cu-based single-atom-alloyed CO adsorption (SAA) dataset~\cite{liang2022multi} 
comprises $3,075$ different Cu-based SAAs adsorbed by CO molecules, with adsorption energies calculated using the DFT. The dataset includes $41$ different element species as dopants. 
By varying the surface index, doping position, and CO molecule adsorption site, $75$ distinct structures are constructed for each doping species. 
These structures are distributed as follows: $6$ on Cu(100), $8$ on Cu(110), $8$ on Cu(111), $27$ on Cu(210), and $26$ on Cu(411) surfaces.

\noindent\textbf{GASpy.} The GASpy dataset~\cite{tran2018active} is related to electrocatalysts for $\text{CO}_2$ reduction and $\text{H}_2$ evolution, including $42,785$ adsorption-energy calculations. Concretely, this dataset considers $1,499$ crystal structures, leading to $17,507$ unique surfaces and $1,684,908$ unique adsorption sites.

\noindent\textbf{Uncertainty Benchmark.} The Uncertainty Benchmark~\cite{tran2020methods} contains $47,279$ DFT-calculated adsorption energies using the Generalized Adsorption Simulator for Python~\cite{tran2018active}. This dataset includes $52$ different elements within the $1,952$ bulk structures for the adsorption surfaces. Additionally, the dataset comprises $9,102 $symmetrically distinct surfaces and $29,843$ distinct coordination environments, defined by the surface and the adsorbate neighbors. The dataset also includes H adsorption energies ($21,269$), CO adsorption energies ($18,437$), OH adsorption energies ($3,464$), O adsorption energies ($2,515$), N adsorption energies ($1,594$).

\noindent\textbf{BioFuels Question Reasoning.} The BioFuels Question Reasoning (BioFuelQR) dataset \citep{sprueill2023monte} is designed to address complex reasoning questions and answers, focusing on the catalysis of the reverse water-gas shift (RWGS) reaction. This is crucial for generating synthetic biofuels with higher selectivity. The questions in this dataset follow a specific template: ``What are the Top-$3$ \{catalyst label\} \{candidate list statement\} that perform the RWGS reaction at a lower temperature ($<200 C$) and demonstrate higher adsorption energy for both CO2 and H2? \{include statement\} \{exclude statement\} Provide scientific explanations and return a list of Top-$3$ answers and their explanations as a list of pairs. The dataset emphasizes the identification of catalysts that can efficiently perform the RWGS reaction at various temperatures while also exhibiting high adsorption energies for both CO2 and H2. It includes detailed reasoning steps, scientific explanations, and a final list of the top three catalysts along with their explanations, presented as pairs.


\subsection{Evaluation Metrics}
Here, we present commonly used evaluation metrics for different tasks in catalyst discovery.

\noindent\textbf{Metrics for Regression Models.} Root Mean Square Error (RMSE), Mean Absolute Error (MAE), and Coefficient of Determination (R2) are commonly used metrics for regression models~\cite{li2020adaptive,zhong2020accelerated}. 


\noindent\textbf{Metrics for Structure to Energy and Forces Task.} Energy MAE, Force MAE, Force cosine, and Energy and Force within Tolerance (EFwT) are four commonly used metrics in the Structure to Energy and Forces (S2EF) task~\cite{shuaibi2022generalizable}.




\noindent\textbf{Effectiveness Metrics for the Initial Structure to Relaxed Structure (IS2RS) Task.} Average Distance within Threshold (ADwT), Force below Threshold (FbT), and Average Force below Threshold (AFbT) are metrics, which are used to evaluate the effectiveness of models in the IS2RS task~\cite{chanussot2021open}.




\noindent\textbf{Other Metrics.} In addition to Energy MAE, \textbf{Energy within a Threshold (EwT)} is also used as a metric in the IS2RE task, calculated as the percentage of computed relaxed energies within $\epsilon = 0.02$ eV of the ground truth relaxed energy.

In the task of catalyst generation, it is usually necessary to consider the metrics of Validity, Uniqueness, and Novelty (i.e., the percentage of samples that are valid, unique, and novel), with their calculation methods depending on the specific generation goals and chemical reactions.

\subsection{Open-source real-world automated systems}
Here, we introduce two open-source automated systems related to catalyst discovery, providing robust, user-friendly tools for researchers.

\noindent \textbf{Open Catalyst Project}~\footnote{https://opencatalystproject.org/} provides datasets, baseline models~\footnote{https://github.com/FAIR-Chem/fairchem}, and a public leaderboard~\footnote{https://opencatalystproject.org/leaderboard.html} for training and evaluating AI models in catalyst discovery. Concretely, the project has released the OC20 and OC22 datasets for training machine learning models, encompassing a total of $1.3$ million in molecular relaxations and over $260$ million in DFT calculations. 

\noindent\textbf{ASE}~\footnote{https://gitlab.com/ase/ase} is an open-source package for setting up, manipulating, running, visualizing, and analyzing atomistic simulations, which can be used in catalyst research.

\noindent \textbf{LCMD}~\footnote{https://www.epfl.ch/labs/lcmd/} is a computational and theoretical laboratory in electronic structure theory, with a focus on method development and conceptual work in the fields of homogeneous catalysis and organic molecular materials. The lab developed NaviCat~\footnote{https://github.com/lcmd-epfl/NaviCat}, a platform that aggregates tools and databases for the digital optimization and discovery of catalysts.

\noindent \textbf{RDKit}~\footnote{https://github.com/rdkit/rdkit} offers a comprehensive suite of tools for cheminformatics, making it an invaluable resource. It supports 2D and 3D molecular operations, enabling users to perform various manipulations and analyses on molecular structures.
Furthermore, RDKit includes powerful tools for generating molecular descriptors, which are essential for machine learning models in cheminformatics, such as AI-empowered catalyst discovery.

\section{FUTURE RESEARCH DIRECTIONS AND OPEN ISSUES}\label{sec:8}
\subsection{Insights of Future Directions}
\subsubsection{General Input \& Output Paradigm.}
 Despite significant progress in discovering highly effective catalysts, there is no general paradigm for catalyst discovery. Currently, researchers employ various input and output formats and then design different AI techniques for different tasks. Here, the input types are diverse, such as sequential data, text data, and graph-based representations, leading to a fragmented approach to catalyst discovery. In the future, researchers should focus on designing a unified input and output paradigm that standardizes both input and output formats for catalyst discovery. Meanwhile, it is better to design the multi-functional learning objective (e.g., considering both adsorption energy and product desorbs), which easily balances the different tasks by tuning the hyperparameters. This unified paradigm would integrate diverse data types, enabling a more holistic and cohesive approach to understanding and predicting catalytic properties. This will not only enhance predictive accuracy but also significantly reduce computational costs by leveraging shared insights and methodologies across different catalyst discovery tasks. Ultimately, the development of a unified learning paradigm holds the promise of accelerating the pace of discovery and deployment of novel catalysts.


\subsubsection{Real-Time Catalyst Discovery and Analysis.}
Existing methods typically overlook two critical aspects: model efficiency and the reuse of intermediate results. On one hand, these methods involve complex model architectures that, while accurate, are computationally intensive and slow during both the training and inference phases. For example, some methods~\cite{yang2023curator,pillai2023interpretable} leverage both DFT calculations and deep learning models for catalyst discovery, leading to substantial computational resources and hindering real-time or large-scale applications. On the other hand, almost all approaches neglect the use of intermediate results, opting instead to start learning from scratch for each new task. This results in significant time and resource overheads, as redundant calculations are performed repeatedly.

In the future, researchers should focus on two main strategies to enhance the efficiency of catalyst discovery. First, reducing model complexity is essential. Simplifying models without sacrificing accuracy can lead to faster inference times and lower computational costs. Techniques such as model pruning~\cite{zhu2017prune}, knowledge distillation~\cite{gou2021knowledge,wang2021knowledge,yim2017gift}, and the use of more efficient algorithms can help achieve this goal. Second, researchers could employ various optimization techniques, such as cache management~\cite{gracioli2015survey,barrios2023service,ji2024lbsc}, to store and reuse intermediate results. By caching intermediate data and results from previous computations, models can avoid redundant calculations and quickly build upon prior knowledge. This approach not only saves time but also improves overall computational efficiency. By focusing on these strategies, researchers can develop more efficient and practical methods for catalyst discovery.

\subsubsection{Comprehensive and Continuously Updated Dataset Corpus for Catalyst Discovery.}
To fully harness the potential of large language models for catalyst discovery, the development of specialized dataset corpora is paramount. However, existing datasets introduced in Section~\ref{sec:9.1} focus either on specific reactions or on non-text information, while the size of these datasets is relatively small. This compromises the performance of LLMs in catalyst discovery. In the future, researchers can focus on constructing expansive, high-quality datasets that include a wide range of catalyst-related information, including experimental results, computational data, detailed reaction mechanisms, and well-characterized experimental procedures. This corpus can be collected and updated from diverse data sources such as scientific literature, patent databases, and proprietary industrial data, ensuring comprehensive coverage of known catalysts and their performance metrics. Additionally, the inclusion of metadata, such as synthesis methods, environmental conditions, and catalyst stability, will enable LLMs to learn relationships between different parameters and catalytic activity. By continuously updating and expanding this corpus with the latest research findings and novel experimental data, we can improve the accuracy and generalization ability of LLM predictions, paving the way for innovative catalyst designs and optimized reaction processes. This holistic and dynamic dataset corpus will be instrumental in driving advancements in catalyst discovery, fostering interdisciplinary collaboration, and accelerating the transition from theoretical insights to practical applications.

\subsubsection{The Physical-Induced Catalyst Discovery based on Machine Learning.} In catalyst discovery, existing methods, especially classical methods, predominantly rely on machine learning techniques to predict the properties of thousands of potential catalysts, subsequently leveraging DFT calculations to provide a theoretical foundation for understanding the electronic and energetic properties of these catalysts. While this combined approach has proven effective, it is computationally expensive and requires significant resources for evaluation. Moving forward, future research should focus on developing a novel framework that integrates machine learning methods with physical principles to enhance the efficiency of catalyst discovery. By incorporating domain-specific physical insights and heuristic rules directly into the machine learning models, researchers can significantly reduce the need for extensive DFT calculations. This approach aims to identify the most discriminative and promising catalysts early in the discovery process, thereby minimizing the computational burden and accelerating the overall workflow. The ultimate goal is to create a streamlined, cost-effective framework that maintains high accuracy in predicting catalytic performance while substantially cutting down on the computational costs of either traditional methods or AI-empowered methods.

\subsubsection{Scalability Optimization for Autonomous Catalyst Discovery.}
Machine learning models have been integrated into the catalyst discovery process to predict the properties of potential catalysts and guide the selection of candidates for experimental validation. These models, including deep learning and generative models like Variational Autoencoders, have shown promise in reducing the number of experiments required. However, they are still heavily dependent on the quality and quantity of existing data, and the integration of ML models with experimental validation often involves manual intervention, which limits scalability.

A promising future direction for catalyst discovery is to develop fully autonomous systems that integrate advanced machine learning techniques with automated experiments. For example, researchers can employ cloud-based infrastructure~\cite{geetha2023dual} that can scale the computational resources required for running complex ML models and managing large datasets, providing the flexibility and scalability to handle the demands of an autonomous discovery system. Additionally, this enables collaboration among researchers by providing access to the autonomous system and its data through cloud platforms, which can accelerate the discovery process by leveraging collective expertise.
These systems can explore vast chemical spaces more efficiently, reduce the time and cost associated with catalyst development, and ultimately lead to the discovery of novel and more effective catalysts for various industrial applications.

\subsection{Limitations of AI}
Encouraged by exciting advances such as antibiotics discovery~\cite{stokes2020deep} and AlphaFold2~\cite{jumper2021highly}, the exploitation of AI has become prevalent and extensively integrated into scientific research.
However, there are ongoing debates on how much, and when AI can make breakthrough scientific discoveries.
Researchers are concerned about the tendency to over-generalise the success of well-carved supervised learning problems, such as protein folding, to science in general~\cite{listgarten2024perpetual}.
Scientists find that the evaluation and interpretation of AI-assisted science are currently below what is typically done by human scientists~\cite{peplow2023robot,leeman2024challenges}, and that determination of novelty and significance cannot be done automatically.
The domain-specific knowledge and insight must be integrated into scientific research, otherwise, the AI-assisted discovery models can be viewed as a compressor of existing data that cannot provide genuine comprehension and understanding as human scientists~\cite{listgarten2024perpetual,messeri2024artificial}.
\cite{krenn2022scientific} highlights the potential of AI in scientific research, but also points out the heavy dependency on the quality and amount of input data and the risk of reinforcing existing biases.

The controversy surrounding AI-driven scientific research mirrors the challenges in catalyst discovery. 
This controversy highlights the limitations of AI models in reasoning and generalization capabilities within complex chemical domains. 
Therefore, researchers should focus on constructing heuristic datasets that encompass novel catalysts and new reasoning relationships, thereby enriching the knowledge base available to AI models. 
To further enhance the chemical reasoning capabilities of AI models, researchers can attempt to design learning paradigms based on chemical reactions or underlying physical principles, such as the reaction-driven prompting design. 
This approach ensures that AI models can better understand and discover new catalysts, contributing to genuine advancements in the field.
Furthermore, AI-empowered catalyst discovery models face several challenges as noted in~\cite{messeri2024artificial}. 
Researchers must effectively communicate epistemic risks to non-experts in an accurate and accessible manner. 
Prioritizing interpretability is essential to make the risks and benefits of these models transparent for scientists. 
Additionally, navigating conflicts of interest is crucial, especially when there are intellectual and financial motivations that might lead to overselling of AI.



\section{Conclusion}\label{sec:9}
Due to the superiority of AI techniques, AI-empowered catalyst discovery has attracted significant attention in both academia and industry for its effectiveness and efficiency compared to traditional methods. 
In this survey, we provide a comprehensive review of the most recent advancements in AI-empowered catalyst discovery. 
We propose a holistic classification scheme to organize existing works, covering both homogeneous and heterogeneous catalyst discovery through direct design and inverse design. 
For each category, we briefly outline the main challenges, detail the representative techniques, and discuss their advantages and limitations. 
Furthermore, we suggest several promising directions for future research from the perspective of computer science and computational chemistry. We hope this survey offers readers a clear understanding of the recent progress and provides insights into future developments.

\bibliographystyle{ACM-Reference-Format}
\bibliography{references}

\end{document}